\documentclass[a4paper,11pt]{article}
\usepackage{jheppub} %
\usepackage{longtable} %
\usepackage{booktabs} %
\usepackage{lineno}
\usepackage{placeins} %
\usepackage{graphicx}%
\usepackage{dcolumn}%
\usepackage{bm}%
\usepackage{amssymb}
\usepackage{amsmath}
\usepackage{xspace}
\usepackage{slashed}
\usepackage{comment}
\usepackage{color}
\usepackage{xcolor}
\usepackage{setspace} %

\newcommand{\pp}           {pp\xspace}

\newcommand{\PbPb}         {\mbox{Pb--Pb}\xspace}

\newcommand{\pPb}          {\mbox{p--Pb}\xspace}

\newcommand{\s}            {\ensuremath{\sqrt{s}}\xspace}
\newcommand{\snn}          {\ensuremath{\sqrt{s_{\mathrm{NN}}}}\xspace}
\newcommand{\pt}           {\ensuremath{p_{\rm T}}\xspace}

\newcommand{\y}         {\ensuremath{y}\xspace}

\newcommand{\dsigmadpt}         {\ensuremath{\mathrm{d}\sigma/\mathrm{d}p_{\rm T}}\xspace}
\newcommand{\dsigmady}         {\ensuremath{\mathrm{d}\sigma/\mathrm{d}y}\xspace}

\newcommand{\RpPb}         {\ensuremath{R_{\rm pPb}}\xspace}

\newcommand{\bbbar}      {\ensuremath{\mathrm{b}\rm\overline{b}}\xspace}

\newcommand{\nineH}        {$\sqrt{s}~=~0.9$~Te\kern-.1emV\xspace}
\newcommand{\seven}        {$\sqrt{s}~=~7$~Te\kern-.1emV\xspace}
\newcommand{\twoH}         {$\sqrt{s}~=~0.2$~Te\kern-.1emV\xspace}
\newcommand{\twosevensix}  {$\sqrt{s}~=~2.76$~Te\kern-.1emV\xspace}
\newcommand{\five}         {$\sqrt{s}~=~5.02$~Te\kern-.1emV\xspace}
\newcommand{\eight}         {$\sqrt{s}~=~8.16$~Te\kern-.1emV\xspace}
\newcommand{\thirteen}         {$\sqrt{s}~=~13$~Te\kern-.1emV\xspace}
\newcommand{\twosevensixnn}{$\sqrt{s_{\mathrm{NN}}}~=~2.76$~Te\kern-.1emV\xspace}
\newcommand{\fivenn}       {$\sqrt{s_{\mathrm{NN}}}~=~5.02$~Te\kern-.1emV\xspace}
\newcommand{\sqrtsNN}      {\ensuremath{\sqrt{s_\mathrm{NN}}}\xspace}

\newcommand{\GeVc}         {Ge\kern-.1emV/$c$\xspace}
\newcommand{\MeVc}         {Me\kern-.1emV/$c$\xspace}
\newcommand{\TeV}          {Te\kern-.1emV\xspace}
\newcommand{\GeV}          {Ge\kern-.1emV\xspace}
\newcommand{\MeV}          {Me\kern-.1emV\xspace}
\newcommand{\GeVmass}      {Ge\kern-.2emV/$c^2$\xspace}
\newcommand{\MeVmass}      {Me\kern-.2emV/$c^2$\xspace}

\newcommand{\ee}           {\ensuremath{\rm{e}^{+}\rm{e}^{-}}\xspace}
\newcommand{\ep}           {\ensuremath{\mathrm{ep}}\xspace}

\newcommand{\dzero}        {\ensuremath{{\rm D}^{0}}\xspace}
\newcommand{\dplus}        {\ensuremath{{\rm D}^{+}}\xspace}

\newcommand{\jpsi}         {\ensuremath{{\rm J}/\psi}\xspace}
\newcommand{\Bzero}        {\ensuremath{\rm B^{0}\xspace}}
\newcommand{\Bplus}        {\ensuremath{\rm B^{+}\xspace}}

\newcommand{\LambdaQCD}    {\ensuremath{ \Lambda_\mathrm{QCD}\xspace}}   
\newcommand{\de}           {\ensuremath{\mathrm{d}}\xspace}

\title{Data-driven measurements of the beauty hadron production in \pPb collisions at \snn = 5.02 and 8.16 TeV}%

\author[a]{Liang Dong}
\author[a]{Tianle Liu}
\author[a]{Guangsheng Li}
\author[b]{Xiaozhi Bai}
\emailAdd{xiaozhi.bai@cern.ch}
\author[c]{Fabrizio Grosa}
\author[a]{Yifei Zhang}
\emailAdd{ephy@ustc.edu.cn}
\author[d]{Jianhui Zhu}

\affiliation[a]{Department of Modern Physics, University of Science and Technology of China, Hefei, China}
\affiliation[b]{Institute of Particle Physics, Central China Normal University (CCNU), Wuhan, China}
\affiliation[c]{European Organization for Nuclear Research, Geneva, Switzerland}
\affiliation[d]{Institute of Modern Physics, Fudan University, Shanghai, China}

\abstract{
Beauty-hadron production in proton--nucleus collisions is a particularly powerful probe of cold nuclear matter (CNM) effects: thanks to the large beauty-quark mass, the production is initiated in early hard scatterings and is calculable in perturbative QCD, while the modifications observed in nuclear collisions directly constrain the nuclear parton distribution functions (nPDFs) in a kinematic region complementary to that accessed by charm. Precisely quantifying these effects is a prerequisite for isolating possible signatures of quark--gluon plasma formation in heavy-ion collisions. In this work, the production of open beauty hadrons in \pPb collisions at $\snn = 5.02$ and $8.16$~\TeV is studied via a data-driven Bayesian unfolding of ALICE and LHCb measurements of non-prompt \dzero and \jpsi mesons. The unfolded \pt- and $y$-differential cross sections are mutually consistent across decay channels and compatible with directly reconstructed beauty-hadron measurements. The nuclear modification factor \RpPb is reported as a function of transverse momentum and rapidity, providing the first unfolding-based result spanning $-4.0 < y < 4.0$ of beauty-hadron \RpPb covering the full rapidity range, with a precision comparable to or better than that of existing data. The results are compared with calculations based on nPDFs and can provide valuable constraints for future global nPDF fits. The total inclusive B hadron production cross section in \pPb collisions is also reported. A weaker CNM suppression is observed for beauty hadrons than for charm hadrons, consistent with the expected mass hierarchy of cold-nuclear-matter effects.
}

\begin{document}
\maketitle
\flushbottom

\section{Introduction}\label{sec:introduction} %

The production of hadrons containing heavy quarks in hadronic collisions is a key testing ground for the strong interaction as described by quantum chromodynamics (QCD). Because the charm- and beauty-quark masses are well above the QCD scale \LambdaQCD, their production is calculable in perturbative QCD~\cite{Nason:1987xz,Collins:1985ue,Mangano:1991jk}.
In proton--proton (\pp) collisions, heavy-flavour hadron production cross sections are computed in a factorisation framework as the convolution of the parton distribution functions (PDFs) of the colliding protons, the partonic cross sections, and fragmentation functions (FFs) describing the hadronisation of the produced quarks. In state-of-the-art calculations, the partonic cross sections are evaluated at next-to-leading order (NLO) in the strong coupling $\alpha_s$ with all-order resummation of next-to-leading logarithms, as implemented in FONLL~\cite{Cacciari:1998it,Cacciari:2001td,Cacciari:2012ny} and GM-VFNS~\cite{Benzke:2019usl,Helenius:2023wkn}, and next-to-next-to-leading-order (NNLO) calculations have recently become available for beauty-quark production~\cite{Catani:2020kkl}. Measurements of charm- and beauty-hadron production by the ALICE, ATLAS, CMS, and LHCb Collaborations in \pp collisions at the LHC provide stringent tests of these calculations~\cite{ATLAS:2011ac,CMS:2011yz,ATLAS:2018gqq,LHCb:2013qoa}.
Fragmentation functions cannot be computed perturbatively and are traditionally parametrised from measurements in \ee and \ep collisions~\cite{Peterson:1982ak}. Recent measurements by the ALICE and LHCb Collaborations have, however, shown that the assumption of universal FFs across collision systems is not justified: the fraction of heavy quarks hadronising into baryons is significantly enhanced in \pp collisions relative to \ee and \ep collisions~\cite{ALICE:2023sgl,ALICE:2023wbx,LHCb:2019fns,LHCb:2023wbo}.

Heavy-flavour hadrons in ultrarelativistic heavy-ion collisions are excellent probes of the colour-deconfined medium---the quark--gluon plasma (QGP)---created in such collisions~\cite{Armesto:2006ph,Andronic:2015wma,Lansberg:2019nhx}. To isolate the phenomena associated with deconfinement, however, it is essential first to quantify the effects arising from the presence of nuclei in the absence of QGP formation, known as cold-nuclear-matter (CNM) effects~\cite{Ferreiro:2012iv,Vogt:2010aa}, which are conventionally studied in \pPb collisions. The PDFs of bound nucleons, for instance, are modified by the nuclear environment as a function of the longitudinal parton momentum fraction $x$, the squared momentum transfer $Q^2$, and the nuclear mass number $A$. This modification is commonly described in a collinear framework using nuclear PDFs (nPDFs)~\cite{Lansberg:2016deg,Eskola:2009uj,Kovarik:2015cma,Eskola:2016oht,Eskola:2020ayu,Kusina:2020lyz}. At low $x$, the colour glass condensate (CGC) effective field theory, which describes the saturation regime of QCD~\cite{Fujii:2006ab,McLerran:1993ni,Ma:2018ekb,Fujii:2018xwm}, provides an alternative description. Multiple scattering of partons in the nucleus can also alter the kinematic distributions of the produced hadrons~\cite{Baier:1996sk,Dokshitzer:2001zm}. Beauty hadrons are particularly valuable probes of these CNM effects: their production proceeds predominantly through gluon--gluon fusion, granting direct sensitivity to the nuclear gluon distribution down to $x \sim 10^{-5}$ at forward rapidity, and the large beauty-quark mass preserves the perturbative calculability of the baseline while giving rise to a characteristic mass hierarchy of nuclear modifications between charm and beauty. In addition to these initial-state effects, final-state phenomena resembling those observed in nucleus--nucleus collisions have recently been reported in high-multiplicity \pPb collisions by the LHC experiments~\cite{ALICE:2022exq,CMS:2018loe}. The nuclear modification factor,
\begin{equation}
\RpPb = \frac{1}{A_{pPb}}\frac{{d}^2\sigma_{pPb}/{d}\pt{d}y}{{d}^2\sigma_{pp}/{d}\pt{d}y},
\end{equation}
where ${d}^2\sigma_{pPb(pp)}/{d}\pt{d}y$ is the \pt- and rapidity-differential production cross section in \pPb (\pp) collisions, is the standard observable used to quantify these nuclear modifications. For hard probes, \RpPb is expected to be unity in the absence of nuclear effects.

Measurements of \RpPb for open and hidden charm and beauty hadrons at $\sqrtsNN = 5.02$ and $8.16$~\TeV have been reported by the ALICE, CMS, and LHCb Collaborations~\cite{ALICE:2016yta,ALICE:2014qnh,CMS:2015sfx,LHCb:2019avm,LHCb:2022dmh}, complemented by measurements of $\Upsilon$ production in \PbPb collisions~\cite{CMS:2017qjw}, of \jpsi mesons from $B$-hadron decays in \pPb collisions~\cite{CMS:2020qva}, and of $\Lambda_b^0$ and $\Lambda_c^+$ baryon production~\cite{LHCb:2016mgk,ALICE:2020wfs}. Earlier measurements at the Tevatron~\cite{Abe:1997dq,Acosta:2003ax} and RHIC~\cite{Adare:2006nq}, and of heavy-flavour decay muons at forward rapidity at the LHC~\cite{ALICE:2012ab}, provided important benchmarks for heavy-flavour production in hadronic collisions. In the beauty sector, the most precise measurements rely on the reconstruction of charm hadrons and charmonia originating from beauty-hadron decays~\cite{ALICE:2024oob,ALICE:2022zig,ALICE:2021lmn,LHCb:2017ygo,ALICE:2017iug}---hereinafter denoted non-prompt \dzero and \jpsi mesons---or on the detection of beauty-hadron decay electrons~\cite{ALICE:2020qpp}. Such measurements do not, however, give direct access to the kinematics of the decaying beauty hadron, and therefore provide only limited information on the parent distributions.
To overcome this limitation, statistical methods such as Bayesian unfolding~\cite{DAGOSTINI1995487,2003RPPh66.1383D,Brenner:2020RooFitUnfold,DAgostini:1994fjx} can be used to infer the parent beauty-hadron kinematic distributions from the measured non-prompt spectra. In this work, the technique developed in Ref.~\cite{Bai:2024pxk} for \pp collisions is applied to \pPb collisions to obtain \pt- and $y$-differential beauty-hadron production cross sections and the nuclear modification factor \RpPb. The inputs are the measurements of non-prompt \dzero mesons at midrapidity ($-0.96 < y < 0.04$) in \pPb collisions at $\sqrtsNN = 5.02$~\TeV~\cite{ALICE:2024oob}, and of non-prompt \jpsi mesons at midrapidity ($-1.37 < y < 0.43$), forward rapidity ($1.5 < y < 4.0$), and backward rapidity ($-5.0 < y < -2.5$) in \pPb collisions at $\sqrtsNN = 5.02$ and $8.16$~\TeV~\cite{ALICE:2022zig,ALICE:2021lmn,LHCb:2017ygo}. The unfolded results provide the first measurement of beauty-hadron \RpPb covering the full rapidity range, and the total \bbbar production cross section is obtained by integrating the unfolded differential cross sections over \pt and $y$.

This paper is organised as follows. Section~\ref{sec:BayesianUnfold} describes the Bayesian unfolding technique and the inputs used for the \pPb analysis. Section~\ref{sec:systematics} discusses the systematic uncertainties. Section~\ref{sec:results} presents the pp reference calculation, the unfolded \pPb cross sections, and the \RpPb results, together with comparisons to nPDF-based model calculations. The conclusions are given in Section~\ref{Conclusions}.

\FloatBarrier

\section{Bayesian unfolding analysis}\label{sec:BayesianUnfold} %

The measurement of open beauty production via semi-inclusive decay channels relies on the reconstruction of non-prompt \dzero and \jpsi mesons from the decays of an admixture of beauty hadrons ($B^0$, $B^\pm$, $B_s^0$, and $\Lambda_b$). In this analysis, a Bayesian unfolding technique is used to infer the parent beauty-hadron kinematic distributions from the measured non-prompt daughter spectra. The full description of the method, including the iterative Bayesian formalism and the propagation of uncertainties, is given in Ref.~\cite{Bai:2024pxk}; here only a concise summary is provided, together with the inputs and response matrices specific to the \pPb analysis.

The conversion of the measured non-prompt \dzero and \jpsi yields into beauty-hadron cross sections requires the inclusive branching fractions of the beauty-hadron admixture into each daughter species. For the \jpsi channel, the inclusive branching fraction $\mathcal{B}(b \rightarrow \jpsi + X) = (1.16 \pm 0.10)\%$ is taken from the PDG~\cite{PDG:2024}. For the \dzero channel, the inclusive probability for a weakly decaying $b$ hadron to produce at least one \dzero or $\overline{\dzero}$ meson is evaluated as
\begin{align}\label{eq:Branchratio}
\mathcal{B}_{b \rightarrow D^0 / \bar{D}^0} &= \mathcal{B}\left(b \rightarrow \bar{D}^0+X\right) + f_d \, \mathcal{B}\left(B^0 \rightarrow D^0+X\right) \nonumber \\
&\quad + f_u \, \mathcal{B}\left(B^{+} \rightarrow D^0+X\right) - \mathcal{B}\left(b \rightarrow D^0 \bar{D}^0+X\right),
\end{align}
where the last term removes the double counting of decays containing both a \dzero and a $\overline{\dzero}$. Using the PDG values $\mathcal{B}(b \rightarrow \bar{D}^0+X)=(58.7\pm2.8)\%$, $\mathcal{B}(B^0 \rightarrow D^0+X)=(8.1\pm1.5)\%$, $\mathcal{B}(B^{+} \rightarrow D^0+X)=(8.6\pm0.7)\%$, $\mathcal{B}(b \rightarrow D^0 \bar{D}^0+X)=(5.1^{+2.0}_{-1.8})\%$~\cite{PDG:2024}, and $f_u = f_d = (36.0\pm0.5)\%$ which are from the LHCb pp measurements~\cite{LHCb:2019fns}, the resulting inclusive branching fraction is $\mathcal{B}_{b \rightarrow D^0 / \bar{D}^0} = (59.6 \pm 3.6)\%$. The uncertainty is obtained by propagating the uncertainties of the individual branching fractions and of the $b$-hadron production fractions in quadrature, assuming no correlations among them.

Since the measured beauty-hadron admixture includes both particles and antiparticles, all four charge combinations, $H_b \rightarrow H_c + X$, $H_b \rightarrow \overline{H}_c + X$, $\overline{H}_b \rightarrow \overline{H}_c + X$, and $\overline{H}_b \rightarrow H_c + X$, are accounted for in the response matrix through the EvtGen simulation. For instance, the \Bplus meson decays predominantly into $\overline{\dzero}$ ($\sim$79\%), with a smaller contribution into \dzero ($\sim$8.6\%), and both topologies are included in the simulation. %

\begin{figure}[!htb]
\begin{center}
\includegraphics[width = 7.5cm]{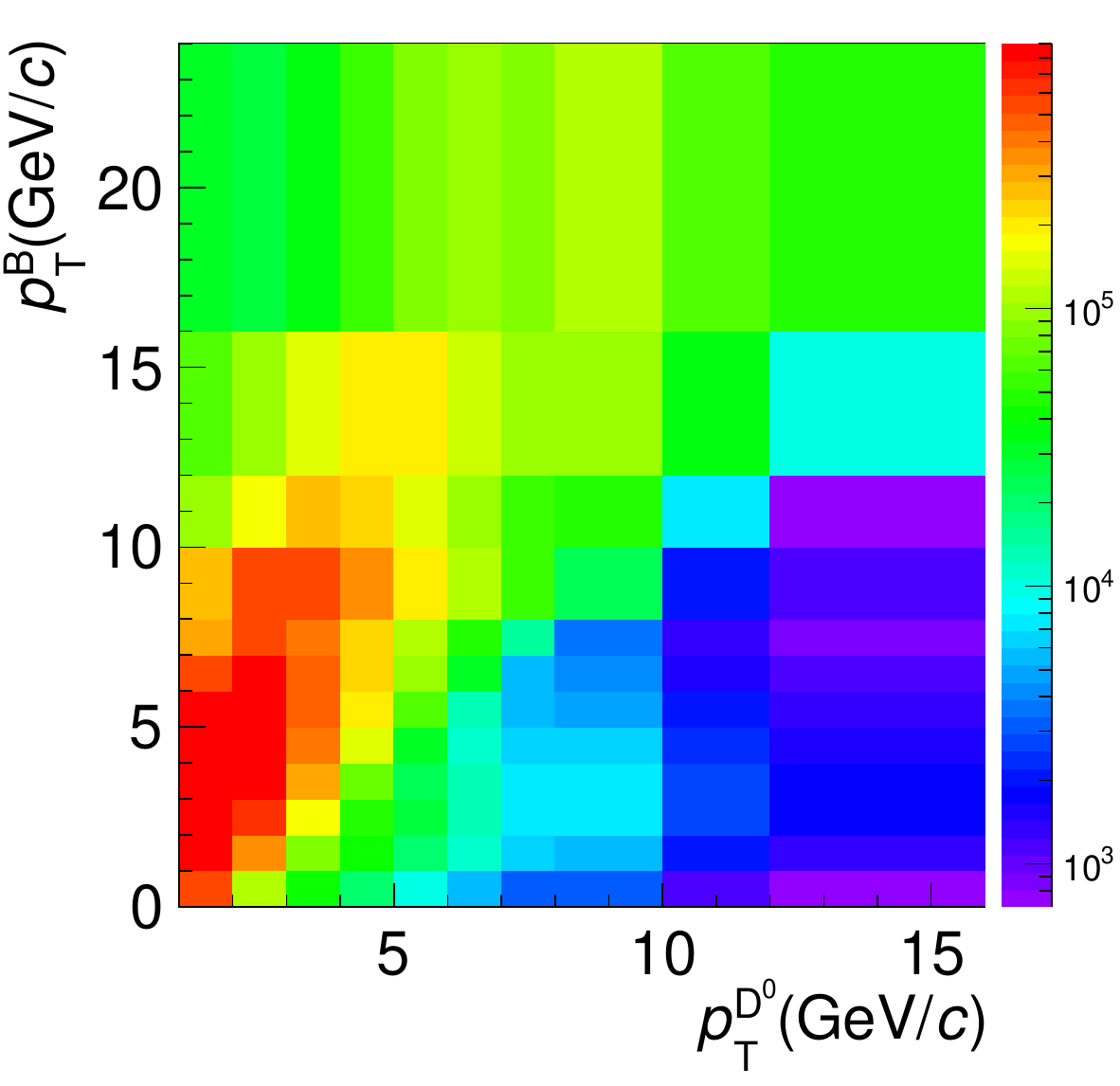}
\includegraphics[width = 7.5cm]{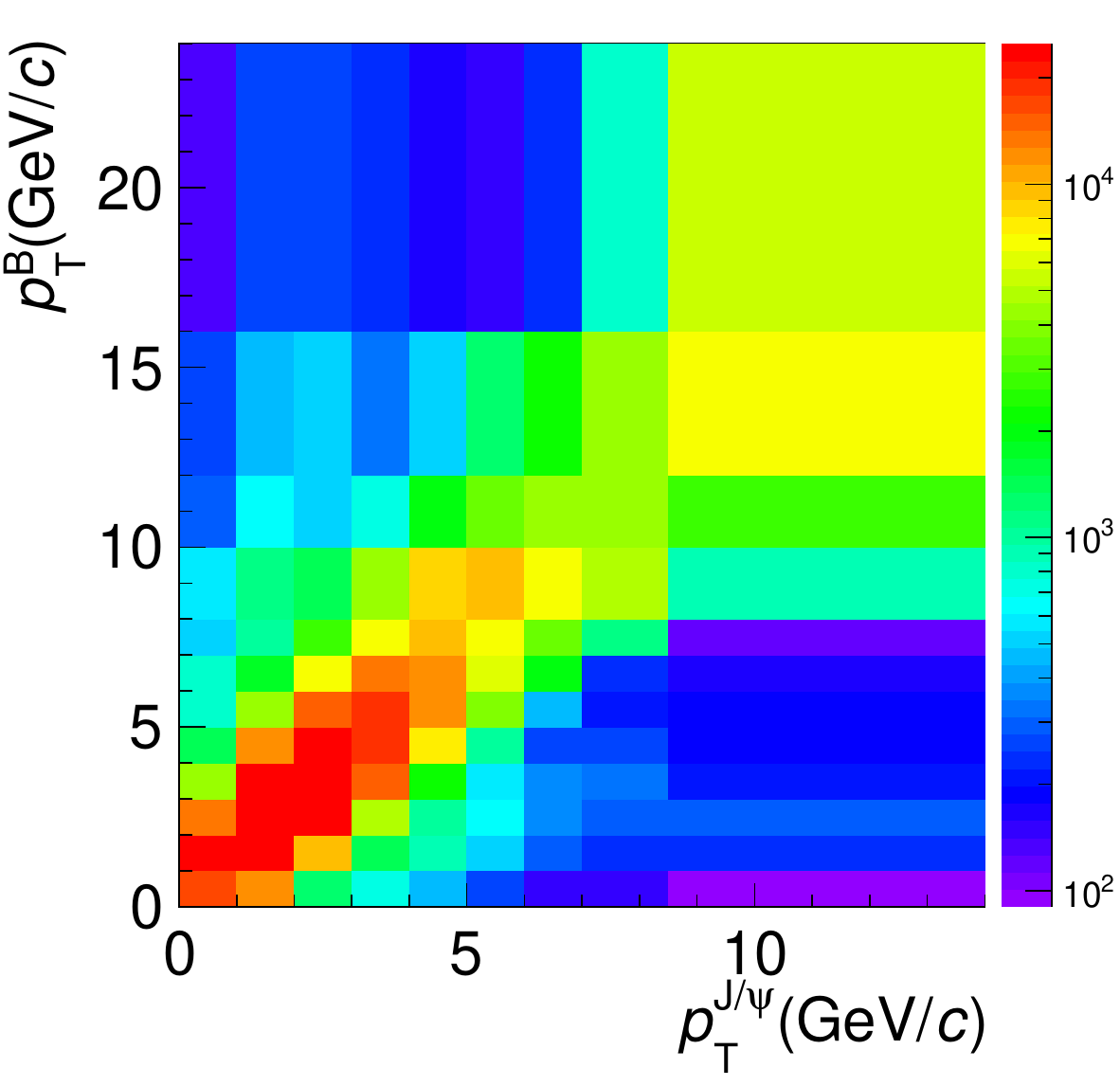}
\caption{Response matrices relating the parent beauty-hadron kinematics to the measured non-prompt \dzero (left panels) and \jpsi (right panels) daughters in \pPb collisions at $\snn = 5.02$~\TeV.}
\label{fig:unford-matrix}
\end{center}
\end{figure}

The correspondence between the parent beauty-hadron kinematics and the measured non-prompt daughter products is encoded in a response matrix, $R_{ij} = P(M_j|E_i)$, derived from EvtGen~2.0~\cite{LANGE2001152} simulations, where $P(M_j|E_i)$ is the probability that a beauty hadron produced in the true kinematic interval $E_i$ yields a daughter measured in the reconstructed interval $M_j$. Final-state QED radiation is simulated with the PHOTOS~3.64 package~\cite{BARBERIO1991115}. The overall efficiency of the kinematic interval $E_i$, including the decay acceptance and branching-fraction effects, is given by the row sum $\epsilon_i = \sum_j P(M_j|E_i)$. The response matrices for the beauty-hadron decays into \dzero and \jpsi in \pPb collisions at $\snn = 5.02$~\TeV are shown in Fig.~\ref{fig:unford-matrix}.

The Bayesian unfolding is performed iteratively, starting from a prior distribution obtained with the PYTHIA~8.309 event generator~\cite{Bierlich:2022pfr}, using the EPPS16nlo nuclear PDF set for the Pb nucleus~\cite{Eskola:2016oht} and the CT14nlo set for the proton~\cite{Dulat:2015mca} through the LHAPDF interface. The iterative procedure follows the D'Agostini prescription~\cite{DAGOSTINI1995487}, and the iteration is stopped when the change between successive iterations becomes smaller than the statistical uncertainties. The cumulative effect of all iterations is included in the uncertainty propagation. Further details on the unfolding formalism are given in Ref.~\cite{Bai:2024pxk}.

\section{Uncertainty analysis}
\label{sec:systematics}

The uncertainties on the unfolded beauty-hadron production cross sections originate from several sources: the statistical and systematic uncertainties of the measured non-prompt \dzero and \jpsi input spectra; the uncertainties on the inclusive beauty-hadron branching fractions into \dzero and \jpsi; the finite size of the EvtGen sample used to build the response matrices; the dependence on the number of Bayesian iterations; the extrapolation of the input spectra to $\pt = 0$; and the choice of the beauty-hadron \pt shape used to construct the response matrix. Each contribution is described in the following and summarised in Table~\ref{tab:sys} for \pPb collisions at $\snn = 5.02$~\TeV and in Table~\ref{tab:sys8p16} for \pPb collisions at $\snn = 8.16$~\TeV.

\subsection{Experimental input uncertainties}

The dominant uncertainties originate from the experimental measurements of the non-prompt \dzero and \jpsi production spectra. The statistical and systematic uncertainties of the ALICE and LHCb inputs are propagated separately to the unfolded results.

The statistical uncertainty on the \pt-integrated cross section is evaluated from the covariance matrix of the input measurement. For a given integration range, all covariance-matrix elements corresponding to the included \pt bins are summed, such that both the diagonal contributions and possible bin-to-bin correlations are retained. The resulting statistical uncertainty is then propagated through the unfolding and the subsequent weighted combination of the different decay channels.

For the experimental systematic uncertainty, the uncertainty in each \pt bin is integrated over the measured spectrum, weighted by the corresponding bin width, and normalized to the \pt-integrated cross section. This prescription corresponds to treating the systematic uncertainties as fully correlated across \pt bins. For the LHCb measurements, the correlated and uncorrelated systematic components provided in each \pt bin are first combined in quadrature, and the resulting uncertainty is then propagated using the same integration procedure.

With this prescription, the input-data statistical and systematic uncertainties on the weighted-average \pt-integrated inclusive B-hadron production cross section in \pPb collisions at $\snn = 5.02$~\TeV are 3.6\% and 3.9\%, respectively. At $\snn = 8.16$~\TeV, the corresponding uncertainties are 1.8\% and 5.9\%, respectively. The fully correlated treatment of the experimental systematic uncertainties is adopted throughout this analysis as a conservative prescription.

\subsection{Branching ratio uncertainties}

The uncertainties on the inclusive branching fractions of beauty hadrons into \dzero and \jpsi are taken from the Particle Data Group (PDG)~\cite{PDG:2024}: $\mathcal{B}(b \rightarrow \jpsi + X) = (1.16 \pm 0.10)\%$ for the \jpsi channel and $\mathcal{B}_{b \rightarrow D^0 / \bar{D}^0} = (59.6 \pm 3.6)\%$ for the \dzero channel, the latter evaluated from the individual $B$-hadron decay branching fractions and $b$-hadron production fractions as described in Section~\ref{sec:BayesianUnfold}. These uncertainties are treated as global (fully correlated) systematic uncertainties. 
The $\Lambda_b^0$-to-$B^0$ production ratio measured by LHCb in \pPb collisions~\cite{LHCb:2019avm} is consistent with the corresponding pp results within experimental uncertainties in the measured kinematic range. Motivated by this observation, the beauty-hadron production fractions inferred from the LHCb pp measurements~\cite{LHCb:2019fns} are adopted as the nominal inputs to Eq.~\ref{eq:Branchratio}, with ($f_u=f_d=(36.0\pm0.5)\%$). No additional enhancement of the $\Lambda_b^0$-to-$B^0$ production ratio relative to pp collisions is assumed.

\subsection{Unfolding-related uncertainties}

The statistical uncertainty of the response matrix originates from the finite size of the EvtGen simulation sample and is reduced by increasing the simulation statistics; it is found to be below 0.3\% for all decay channels and collision systems considered.

The dependence on the number of Bayesian iterations is assessed by comparing the unfolded results obtained with four and six iterations; the difference is below 1\% in all cases and is assigned as a systematic uncertainty.

The choice of the beauty-hadron \pt shape used to construct the response matrix introduces a further uncertainty. To evaluate this effect, an alternative response matrix is constructed by reweighting the generated beauty-hadron \pt distribution using \pt-dependent weights derived from FONLL calculations\cite{Cacciari:2015fta}. The weights are applied according to the transverse momentum of the parent beauty hadron when filling the response matrix. The data are then unfolded using the nominal and reweighted response matrices, and the difference between the resulting cross sections is assigned as the systematic uncertainty associated with the beauty-hadron \pt shape. The resulting uncertainty is below 1\% for all channels, except for the ALICE non-prompt \jpsi channel at $\snn = 8.16$~\TeV, where it reaches 1.5\%.

\subsection{Extrapolation uncertainty}

The ALICE measurements of non-prompt \dzero and \jpsi extend down to $\pt = 1$~\GeVc, while the LHCb non-prompt \jpsi measurement reaches $\pt = 0$. In principle, the unfolding procedure recovers the unmeasured low-\pt region without any extrapolation; nevertheless, the stability of the results is tested by comparing the unfolded spectra obtained with and without an explicit extrapolation of the input data to $\pt = 0$. The difference amounts to about 1\% for the ALICE non-prompt \dzero channel at $\snn = 5.02$~\TeV and about 11\% for the ALICE non-prompt \jpsi channel at $\snn = 8.16$~\TeV, and is assigned as a systematic uncertainty.

\subsection{Total uncertainties}

The correlated and uncorrelated contributions from each source are propagated separately, and the statistical uncertainties of the final results are those of the input measurements. All systematic contributions are added in quadrature to obtain the total systematic uncertainty.

The total uncertainty on the \pt-integrated inclusive B hadron production cross section at midrapidity, obtained from the weighted average of the individual channels, is 7.3\% for \pPb collisions at $\snn = 5.02$~\TeV and 10.7\% at $\snn = 8.16$~\TeV; the individual contributions are reported in Table~\ref{tab:sys} and Table~\ref{tab:sys8p16}, respectively. The larger total uncertainty of the ALICE non-prompt \jpsi channel at 8.16~\TeV (20.1\%) is driven by the combination of limited statistics and the sizable extrapolation uncertainty; this channel enters the weighted average through its unfolded spectrum rather than as an independent \pt-integrated measurement.

\begin{table*}[!htbp]
\centering
\caption{Relative systematic uncertainties (in percent) on the \pt-integrated inclusive B hadron production cross section at midrapidity in \pPb collisions at $\snn = 5.02$~\TeV. The systematic sources are treated as fully correlated across \pt bins. The total is obtained by adding the individual contributions in quadrature.}
\label{tab:sys}
\begin{tabular*}{\textwidth}{@{\extracolsep{\fill}}lcccc}
\toprule
Source & \dzero (ALICE) & \jpsi (ALICE) & \jpsi (LHCb) & Weighted avg. \\
\midrule
Input data statistical & 7.3 & 15.2 & 2.5 & 3.6 \\
Input data systematic & 7.2 & 11.0 & 5.0 & 3.9 \\
Response matrix stat. & $<$0.1 & 0.1 & 0.2 & 0.1 \\
Branching ratio & 6.0 & 8.6 & 8.6 & 4.9 \\
Bayesian iterations & 0.5 & 0.3 & $<$0.1 & 0.2 \\
Low-$\pt$ extrapolation & 2.6 & 2.2 & --- & 1.0 \\
Beauty-hadron $\pt$ shape & 0.1 & 0.5 & 0.4 & 0.2 \\
\midrule
Total uncertainty & 12.1 & 20.8 & 10.3 & 7.3 \\
\bottomrule
\end{tabular*}
\end{table*}

\begin{table*}[!htbp]
\centering
\caption{Relative systematic uncertainties (in percent) on the \pt-integrated inclusive B hadron production cross section at midrapidity in \pPb collisions at $\snn = 8.16$~\TeV. The systematic sources are treated as fully correlated across \pt bins. The total is obtained by adding the individual contributions in quadrature. }
\label{tab:sys8p16}
\begin{tabular*}{\textwidth}{@{\extracolsep{\fill}}lccc}
\toprule
Source & \jpsi (ALICE) & \jpsi (LHCb) & Weighted avg. \\
\midrule
Input data statistical & 11.7 & 0.5 & 1.8 \\
Input data systematic & 8.1 & 7.2 & 5.9 \\
Response matrix stat. & 0.3 & 0.2 & 0.2 \\
Branching ratio & 8.6 & 8.6 & 8.6 \\
Bayesian iterations & 0.8 & 0.3 & 0.3 \\
Low-$\pt$ extrapolation & 11.2 & --- & 1.4 \\
Beauty-hadron $\pt$ shape & 1.5 & 0.5 & 0.5 \\
\midrule
Total uncertainty & 20.1& 11.2& 10.7  \\
\bottomrule
\end{tabular*}
\end{table*}

\FloatBarrier

\section{Results}\label{sec:results} %

In this section, \pt-differential (\dsigmadpt) and \y-differential (\dsigmady, 0 $<$ \pt $<$ 24 \GeVc) cross sections of \bbbar production in \pPb collisions at \snn = 5.02 and 8.16~\TeV are presented. These results are obtained with the Bayesian unfolding procedure described in Section~\ref{sec:BayesianUnfold}, applied to the measured non-prompt \dzero and \jpsi spectra.

The weighted-average \pt-differential \bbbar production cross section is computed from the unfolded results of the different decay channels (Eqs.~\ref{eq:averageD} and \ref{eq:averageDuncer}), with weights given by the inverse of the squared total uncertainties, so that channels with smaller uncertainties carry a larger weight. This procedure, previously employed by ALICE for averaging charm-meson yields in \PbPb collisions at \fivenn~\cite{ALICE:2021rxa}, is applied here to combine the non-prompt \dzero and \jpsi channels. 
 
\begin{equation}\label{eq:averageD}
\frac{{d}\sigma_{{ave.}}}{{d}\pt} = \sum_{i} w_{i} \, \frac{{d}\sigma_{i}}{{d}\pt}
\end{equation}

\begin{equation}\label{eq:averageDuncer}
w_{i} = \frac{1/\sigma_{i}^{2}}{\sum_{j} 1/\sigma_{j}^{2}}
\end{equation}

\subsection{Reference open beauty production cross section in \pp collisions at \s = 8.16 TeV}

The calculation of \RpPb requires a \pp reference cross section at the same centre-of-mass energy as the \pPb measurement. Since no \pp data were collected at \s = 8.16~\TeV, the reference is obtained by extrapolating the unfolded \pp results at \s = 5.02 and 13~\TeV~\cite{Bai:2024pxk}---based on the ALICE non-prompt \dzero~\cite{ALICE:2021mgk,ALICE:2024xln} and LHCb non-prompt \jpsi~\cite{LHCb:2021pyk,LHCb:2015foc} measurements in \pp collisions---using as a scaling factor the ratio of FONLL predictions at the two energies, using as a scaling factor the ratio of FONLL predictions at the two energies~\cite{LHCb:2017vec}:

\begin{equation}
(\de\sigma/\de\pt)_{scaled}^{8.16~{\TeV}} = \frac{(\de\sigma/\de\pt)_{FONLL}^{8.16~{\TeV}}}{(\de\sigma/\de\pt)_{FONLL}^{5.02\text{ or }13~{\TeV}}}\cdot(\de\sigma/\de\pt)_{data}^{5.02\text{ or }13~{\TeV}}
\label{eq:ppscale}
\end{equation}

Using Eq.~\ref{eq:ppscale}, the open beauty production cross section in \pp collisions at \s = 8.16~\TeV is obtained independently from the unfolded results at \s = 5.02 and 13~\TeV.

The two \pp references at \s = 8.16~\TeV, extrapolated independently from the unfolded results at \s = 5.02 and 13~\TeV, are both retained in the analysis. Their mutual consistency provides a cross-check of the energy extrapolation and of the open beauty production cross sections obtained from the two unfolded datasets~\cite{Bai:2024pxk}.

\begin{figure}[!htb]
 	\begin{center}
     		\includegraphics[width = 7.5cm]{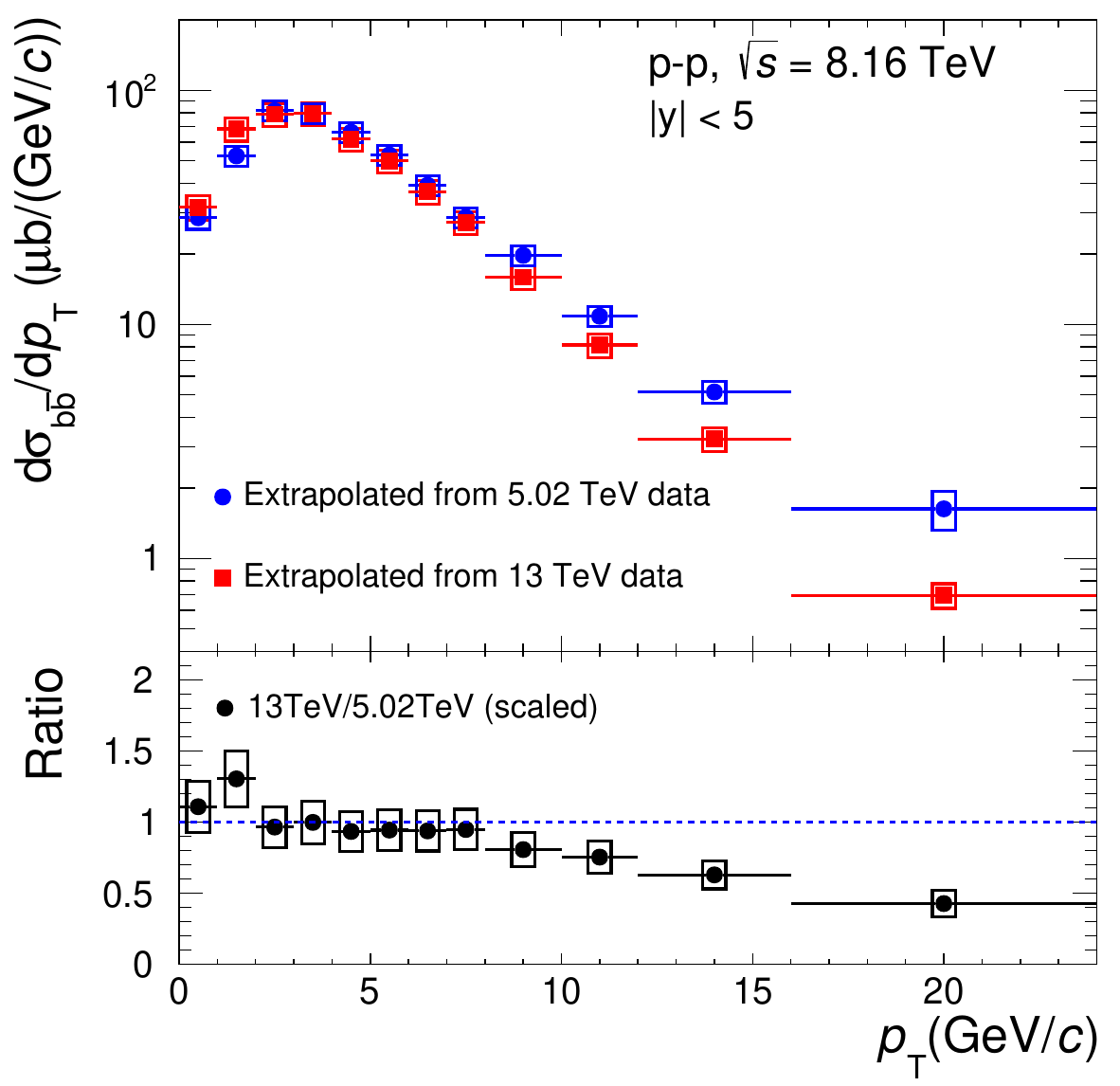}
 		\includegraphics[width = 7.5cm]{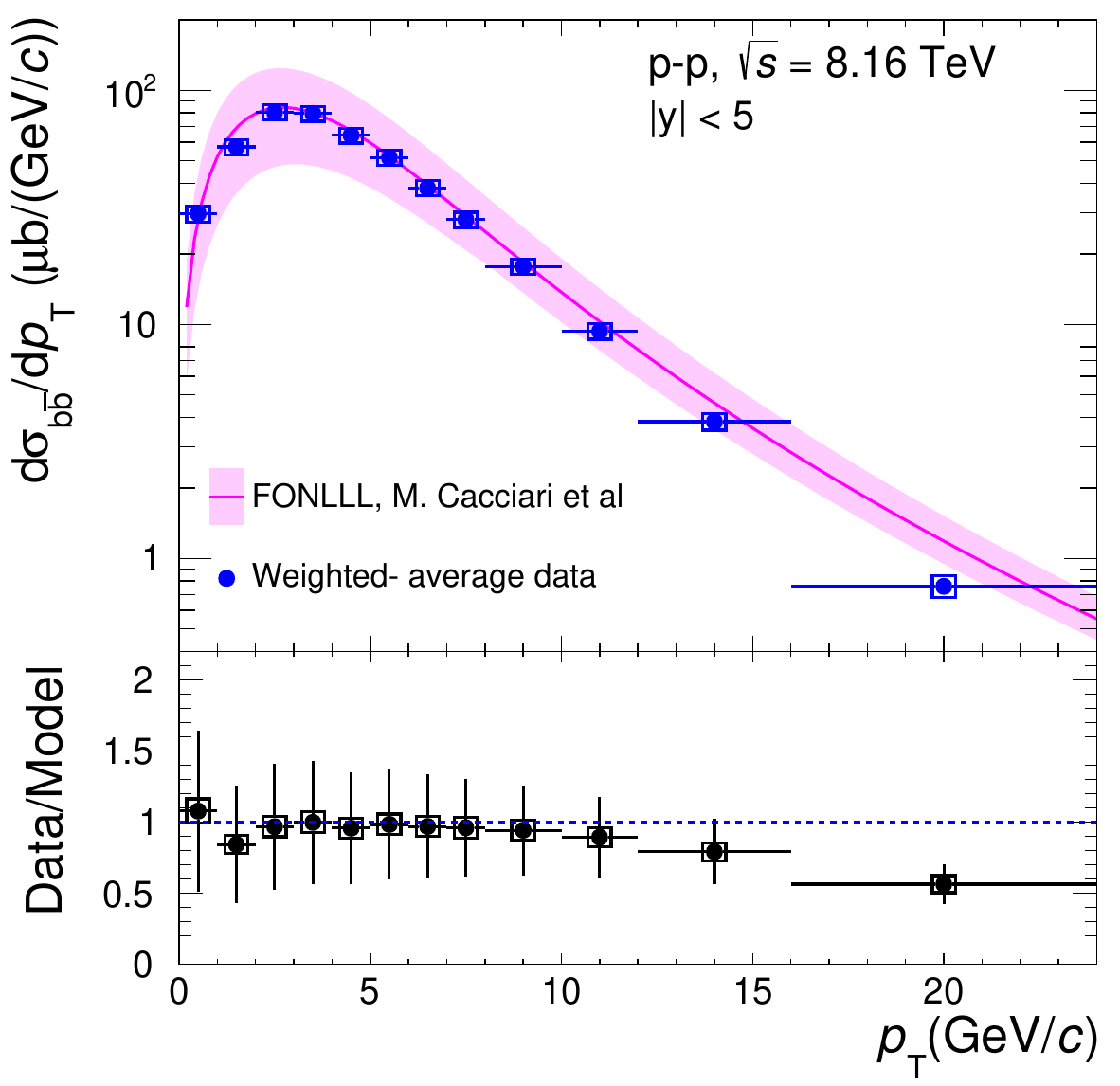}
 		\caption{\pt-differential pp reference cross sections for $|y| < 5$. Left: extrapolated results from 5.02~\TeV (blue) and 13~\TeV (red) data with their ratio in the lower panel. Right: weighted-average result at \s = 8.16~\TeV (blue) compared with FONLL predictions (magenta band) with their ratio in the lower panel.}
 		\label{fig:ppy10ref}
 	\end{center}
 \end{figure}

\begin{figure}[!htb]
 	\begin{center}
     		\includegraphics[width = 7.5cm]{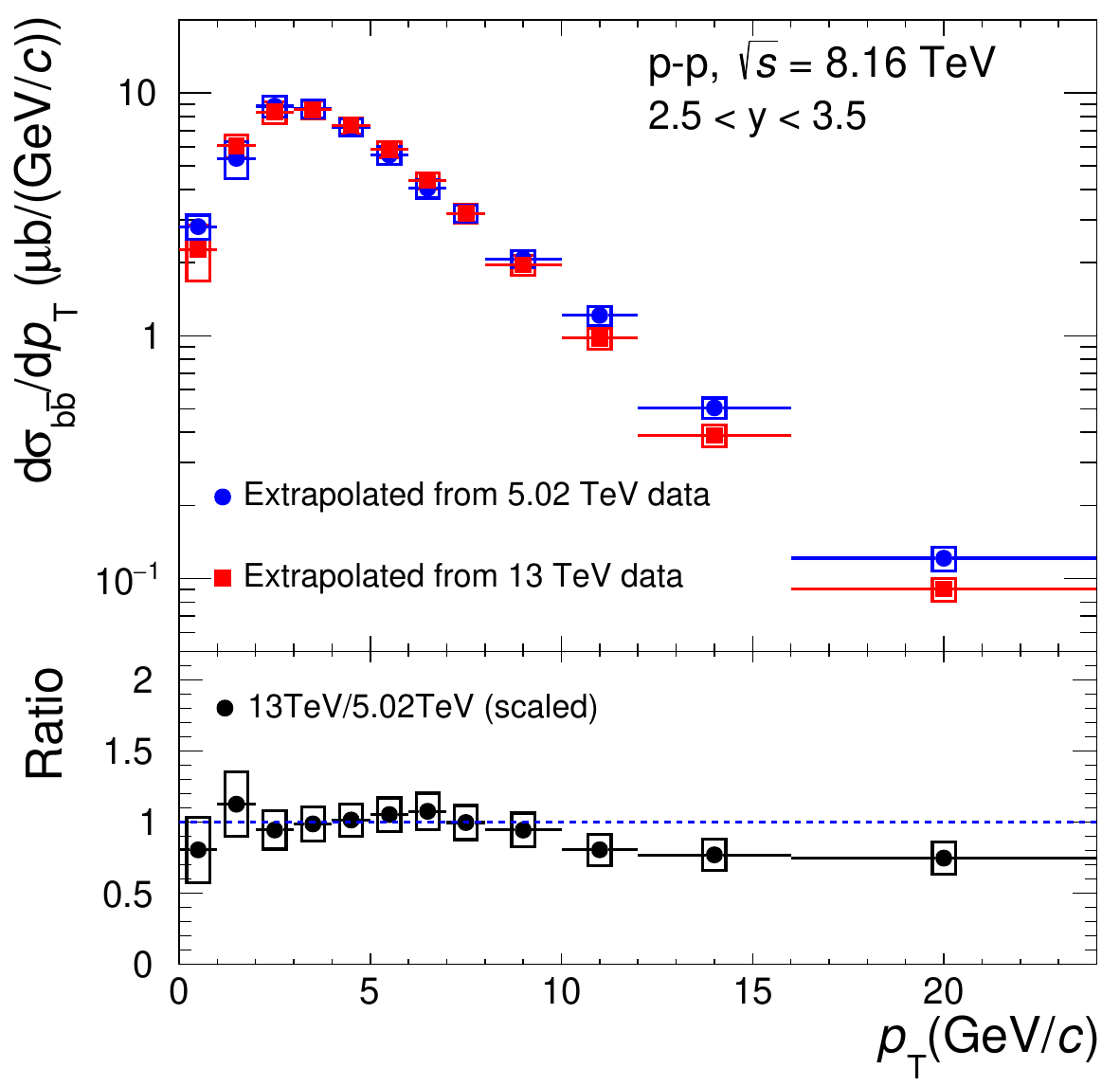}
 		\includegraphics[width = 7.5cm]{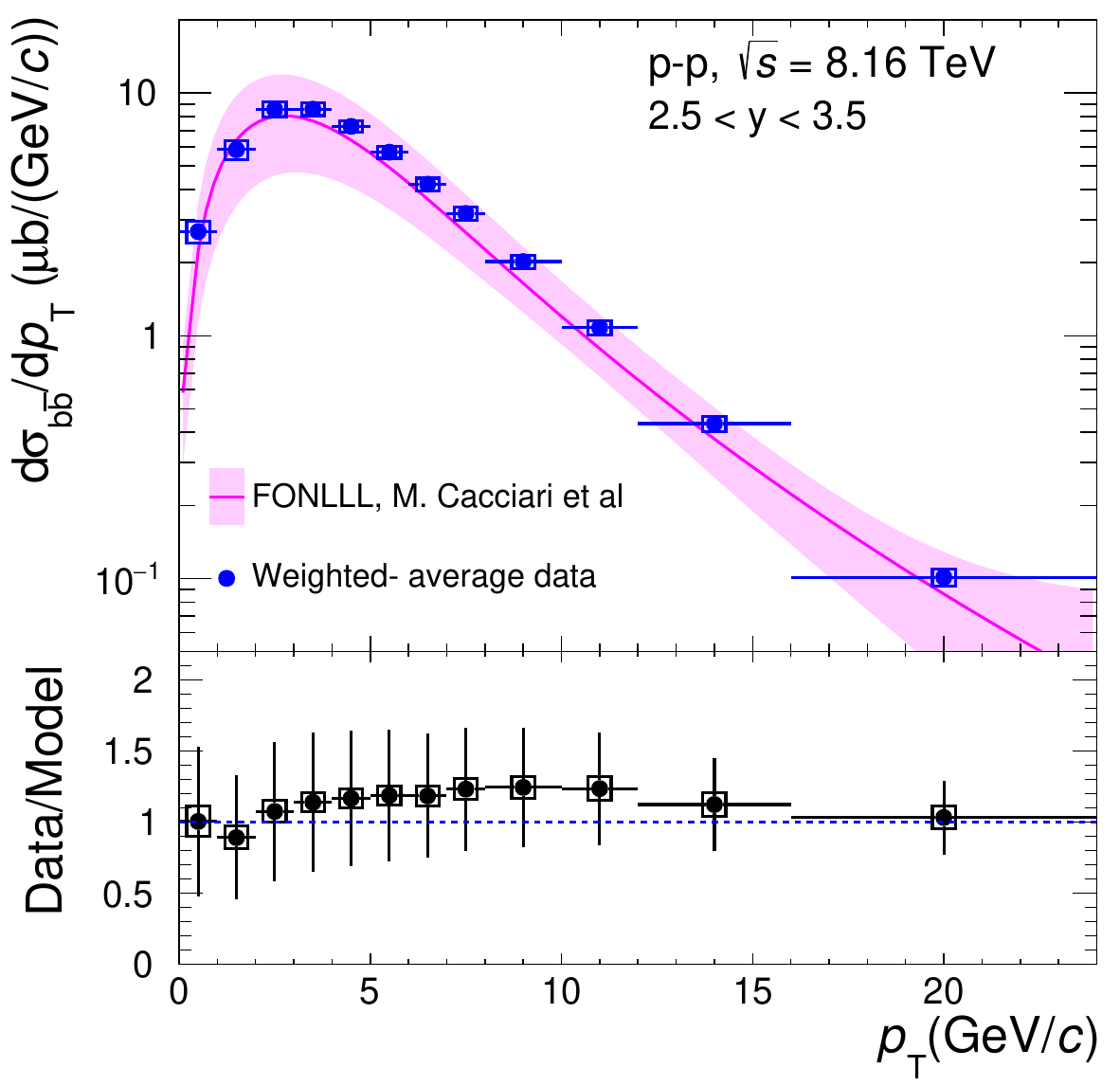}
 		\caption{\pt-differential pp reference cross sections for $2.5 < y < 3.5$. Left: extrapolated results from 5.02~\TeV (blue) and 13~\TeV (red) data with their ratio in the lower panel. Right: weighted-average result at \s = 8.16~\TeV (blue) compared with FONLL predictions (magenta band) with their ratio in the lower panel.}
 		\label{fig:ppfwdref}
 	\end{center}
 \end{figure}

\begin{figure}[!htb]
 	\begin{center}
     		\includegraphics[width = 7.5cm]{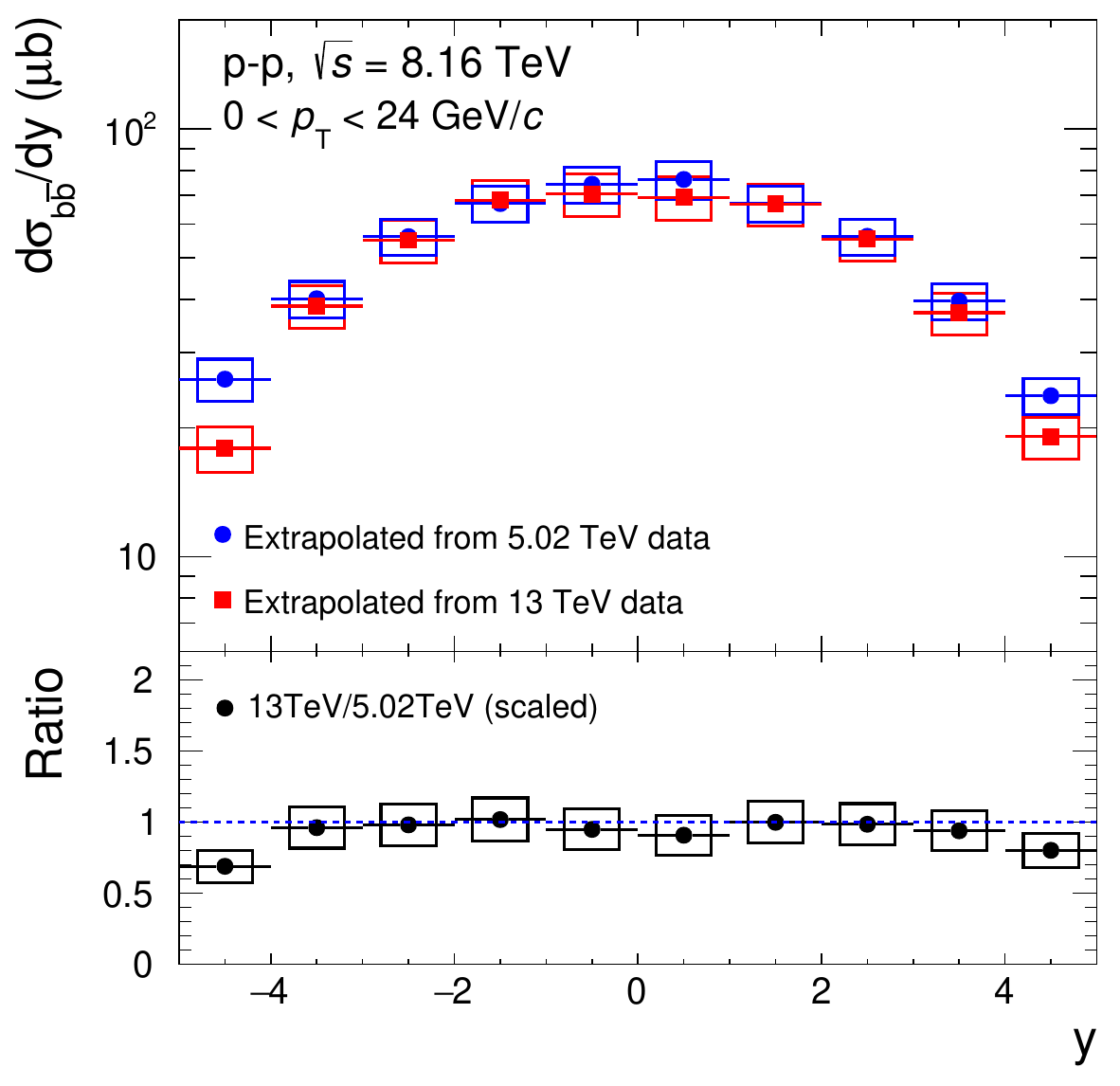}
 		\includegraphics[width = 7.5cm]{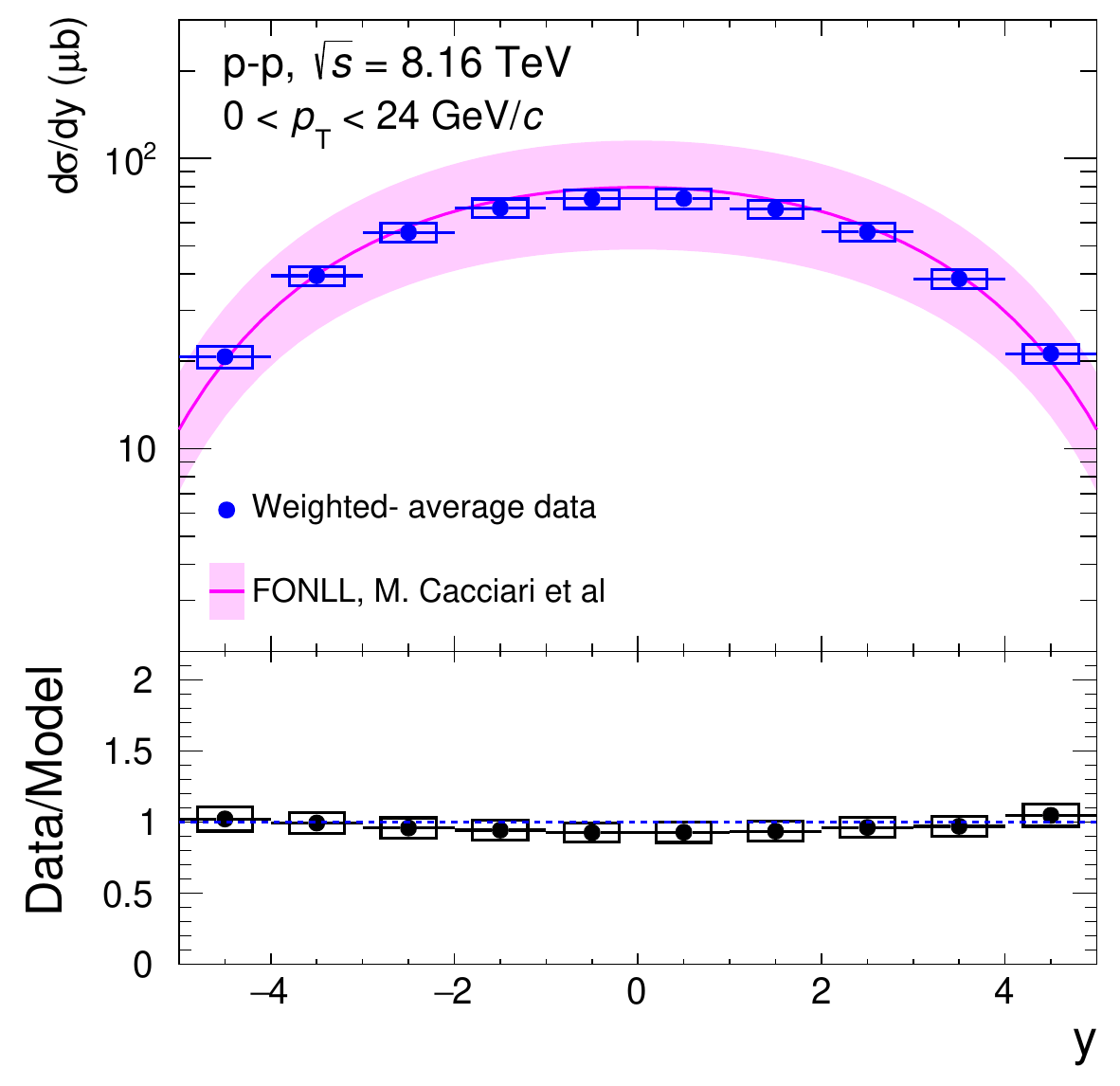}
 		\caption{\y-differential pp reference cross sections for $\pt < 24$~\GeVc. Left: extrapolated results from 5.02~\TeV (blue) and 13~\TeV (red) data with their ratio in the lower panel. Right: weighted-average result at \s = 8.16~\TeV (blue) compared with FONLL predictions (magenta band) with their ratio in the lower panel.}
 		\label{fig:ppyref}
 	\end{center}
 \end{figure}

Statistical uncertainties are propagated directly through the scaling procedure. The FONLL extrapolation uncertainties account for variations of the renormalisation and factorisation scales, the beauty-quark mass, and the PDFs~\cite{Cacciari:2012ny}; the upper and lower bounds in each bin are taken as the maximum and minimum ratios among all theoretical variations. The total systematic uncertainty is obtained by adding the scaled experimental systematic uncertainties and the FONLL ratio uncertainties in quadrature. The two extrapolated \pt-differential \pp cross sections---one from the 5.02~\TeV data and one from the 13~\TeV data---are then combined using inverse-variance weighting (Eqs.~\ref{eq:averageD}--\ref{eq:averageDuncer}) to give the final \pp reference at \s = 8.16~\TeV. The resulting \pt- and \y-differential cross sections are shown in Figs.~\ref{fig:ppy10ref}--\ref{fig:ppyref}.

 The two extrapolated \pp references are mutually consistent within uncertainties over the full \pt and rapidity ranges, and their weighted average agrees with the FONLL predictions within uncertainties. This validates the extrapolation procedure and justifies the use of the weighted-average result as the \pp reference for the \RpPb calculation.

\subsection{Open beauty production cross sections in \pPb collisions at \snn = 5.02 and 8.16~\TeV}

\begin{figure}[!htb]
 	\begin{center}
 		\includegraphics[width = 7.5cm]{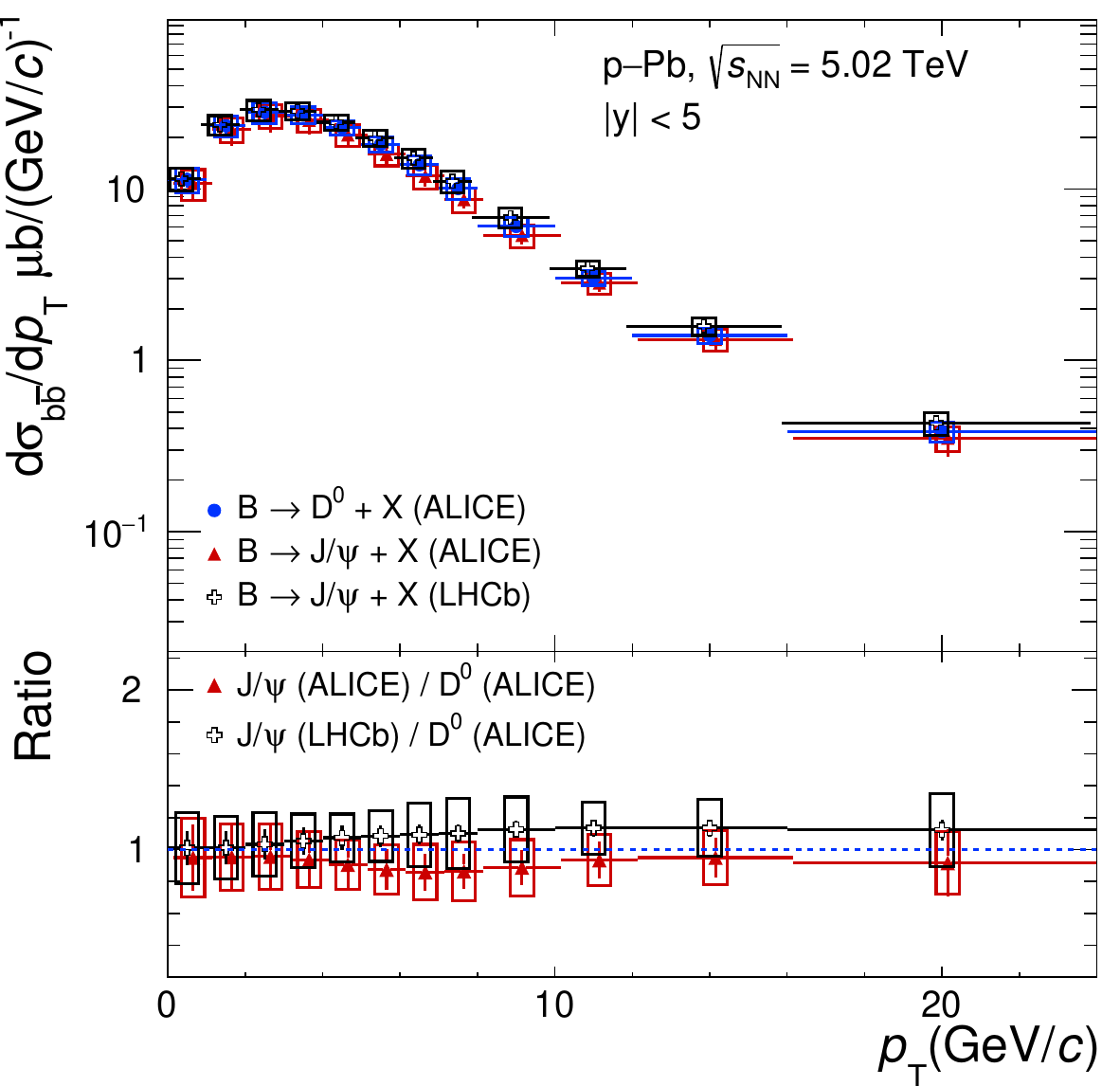}
        \includegraphics[width = 7.5cm]{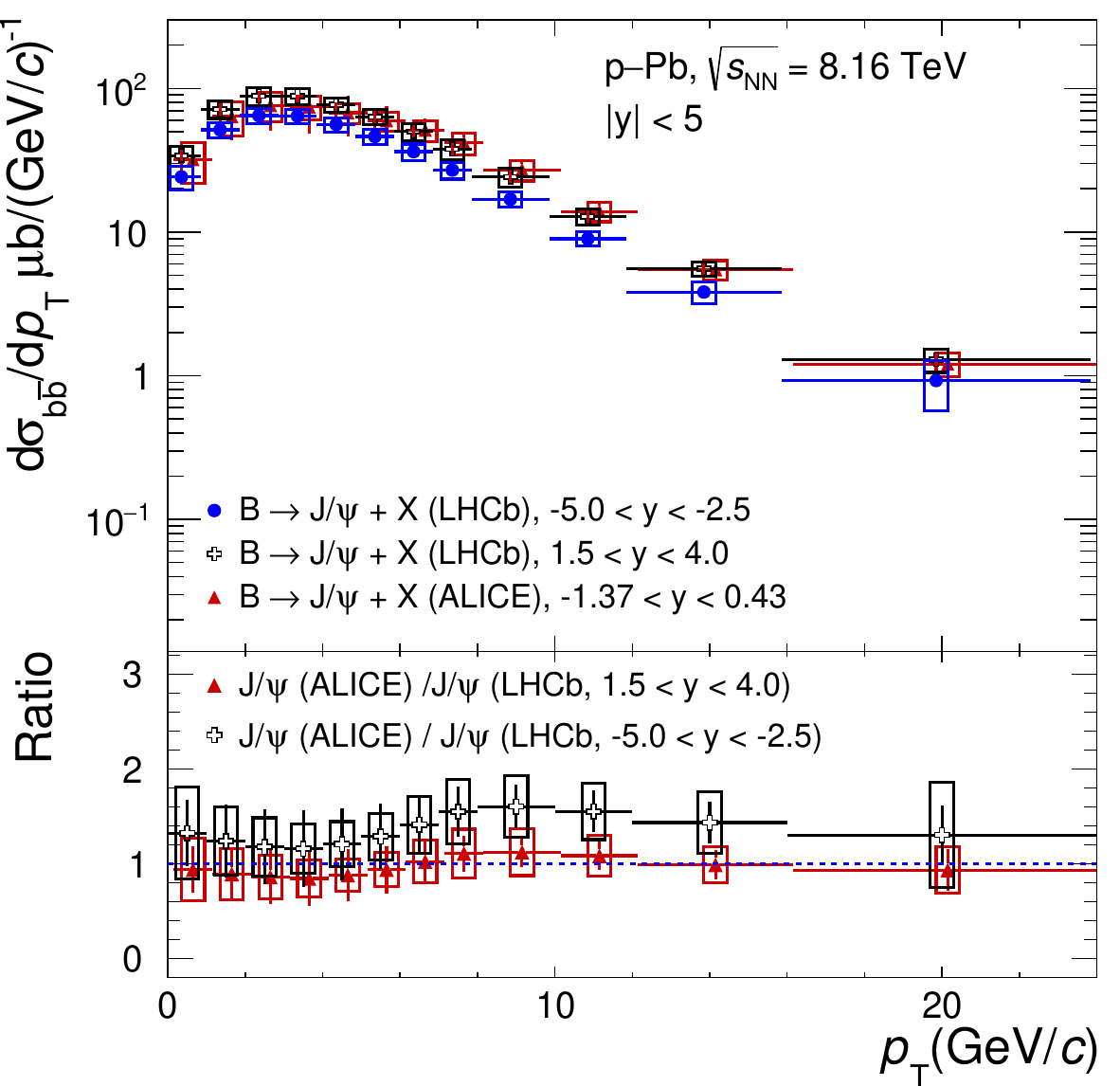}
 		\caption{Unfolded \bbbar production cross section \dsigmadpt as a function of \pt in \pPb collisions at $|y| < 5 $. Left: \snn = 5.02~\TeV results from non-prompt \dzero (blue) and \jpsi (red, black) decay channels. Right: \snn = 8.16~\TeV results from non-prompt \jpsi measured by ALICE (red) and LHCb (blue, black).}
 		\label{fig:PtSpectra_three_5020-ratio}
 	\end{center}
 \end{figure}

The unfolded \pt-differential cross sections \dsigmadpt are shown in Fig.~\ref{fig:PtSpectra_three_5020-ratio} for $|y| < 5$. The inputs to the unfolding are the measured non-prompt \dzero spectra from ALICE~\cite{ALICE:2024oob}, the non-prompt \jpsi spectra from ALICE~\cite{ALICE:2021lmn} and non-prompt J/$\psi$ spectra from LHCb~\cite{LHCb:2013gmv} for \pPb collisions at \snn = 5.02~\TeV, and the non-prompt \jpsi spectra from ALICE and LHCb~\cite{ALICE:2022zig,LHCb:2017ygo} for \pPb collisions at \snn = 8.16~\TeV. In the left panel, the unfolded results from the non-prompt \dzero (blue) and non-prompt \jpsi (red and black) decay channels at \snn = 5.02~\TeV are compared; in the right panel, the results from the non-prompt \jpsi measurements by ALICE (red) and LHCb (blue and black) at \snn = 8.16~\TeV are shown, with the ratios between the different channels displayed in the lower panels. The results from the two semi-inclusive decay channels are consistent within uncertainties once unfolded to the same rapidity interval ($|y| < 5$), despite the very different rapidity coverages of the original measurements: the non-prompt \dzero is measured at midrapidity ($-0.96 < y < 0.04$), whereas the non-prompt \jpsi measurements cover midrapidity ($-1.37 < y < 0.43$) and forward rapidity ($1.5 < y < 4.0$). This cross-channel consistency provides a cross-check that the decay products are mapped back to a common underlying beauty-hadron distribution and supports the response matrices obtained from the EvtGen simulation.

\begin{figure}[!htb]
    \begin{center}
     	\includegraphics[width = 7.5cm]{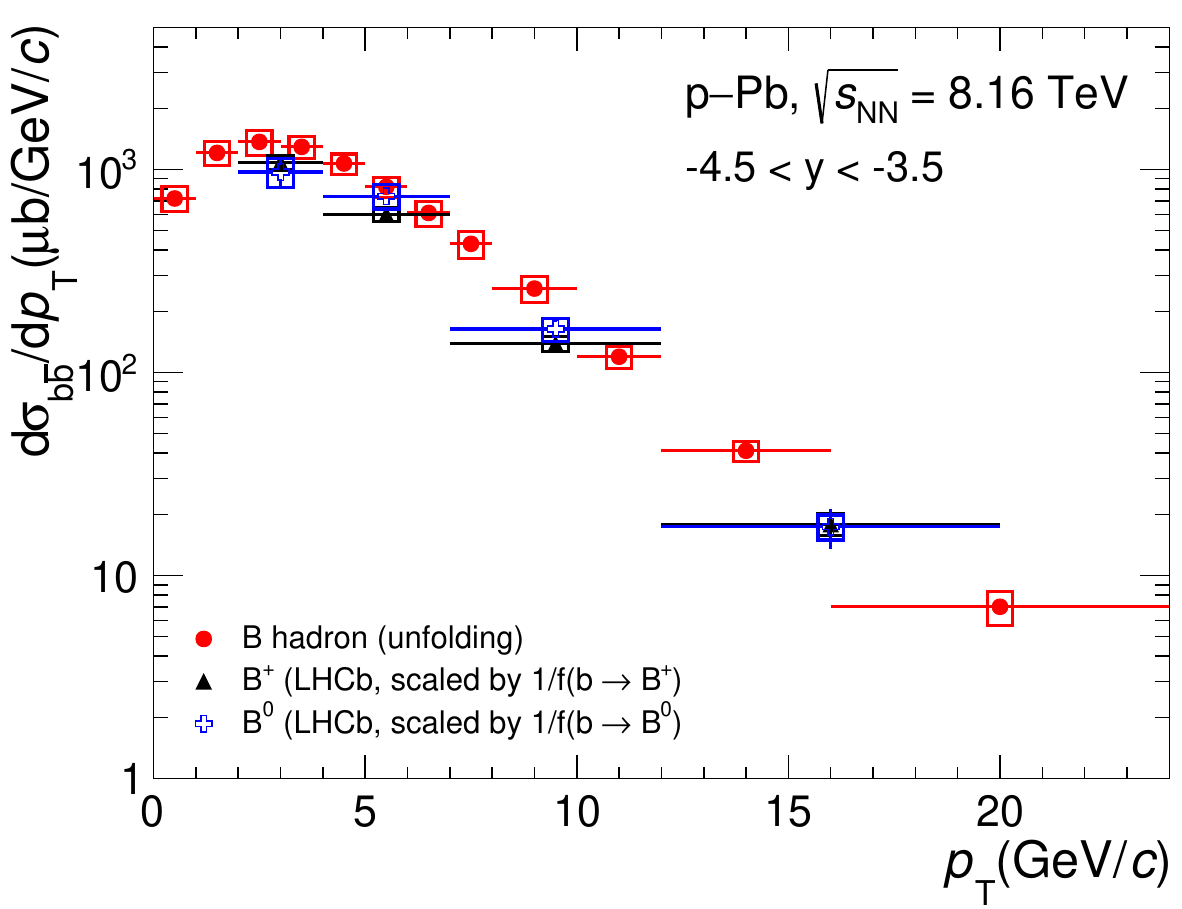}
 		\includegraphics[width = 7.5cm]{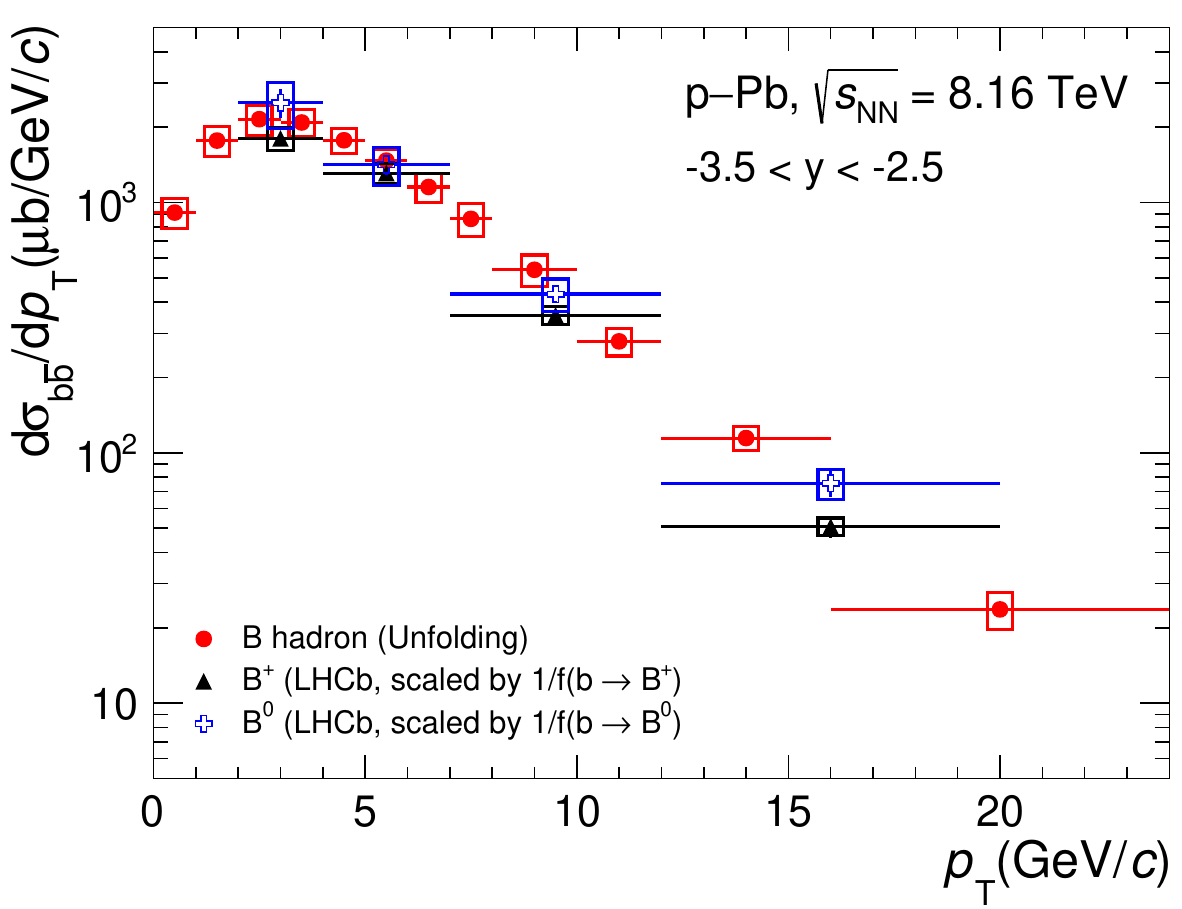}
            \includegraphics[width = 7.5cm]{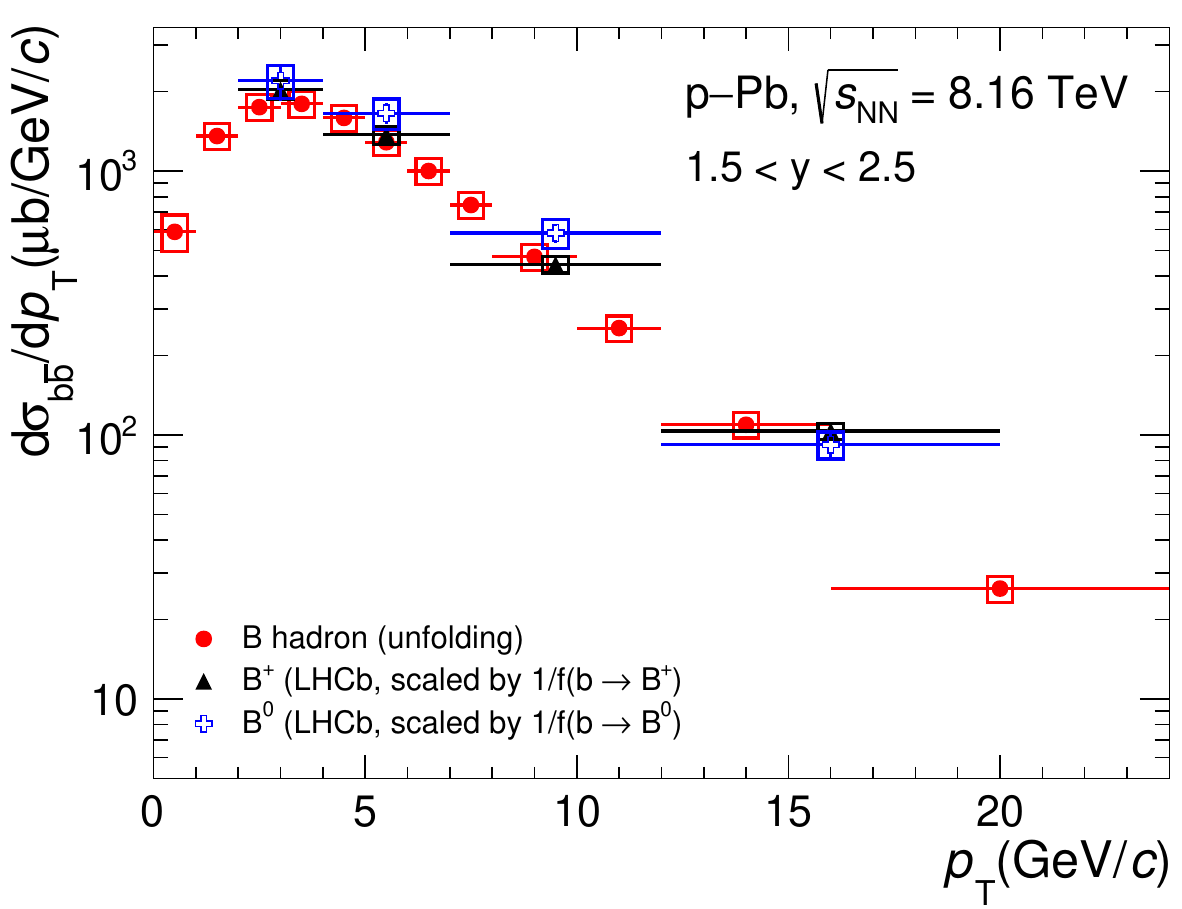}
 		\includegraphics[width = 7.5cm]{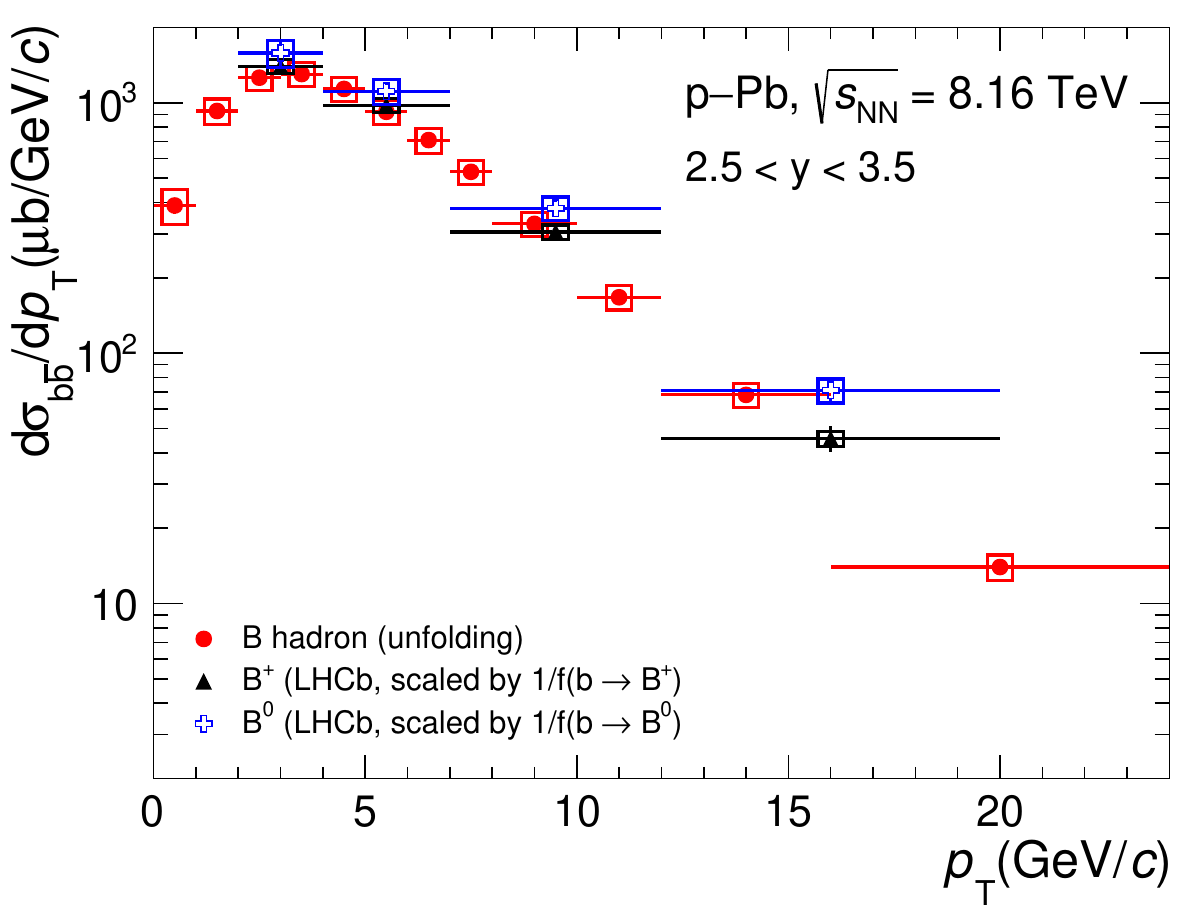}
 		\caption{\pt-differential unfolded results (red) scaled by production fractions and compared with LHCb measurements of \Bplus (black) and \Bzero (blue) in the rapidity intervals $(-4.5,-3.5)$, $(-3.5,-2.5)$, $(1.5,2.5)$, and $(2.5,3.5)$\cite{LHCb:2019avm}.}
 		\label{fig:exppt}
 	\end{center}
 \end{figure}

\begin{figure}[!htb]
    \begin{center}
     	\includegraphics[width = 7.5cm]{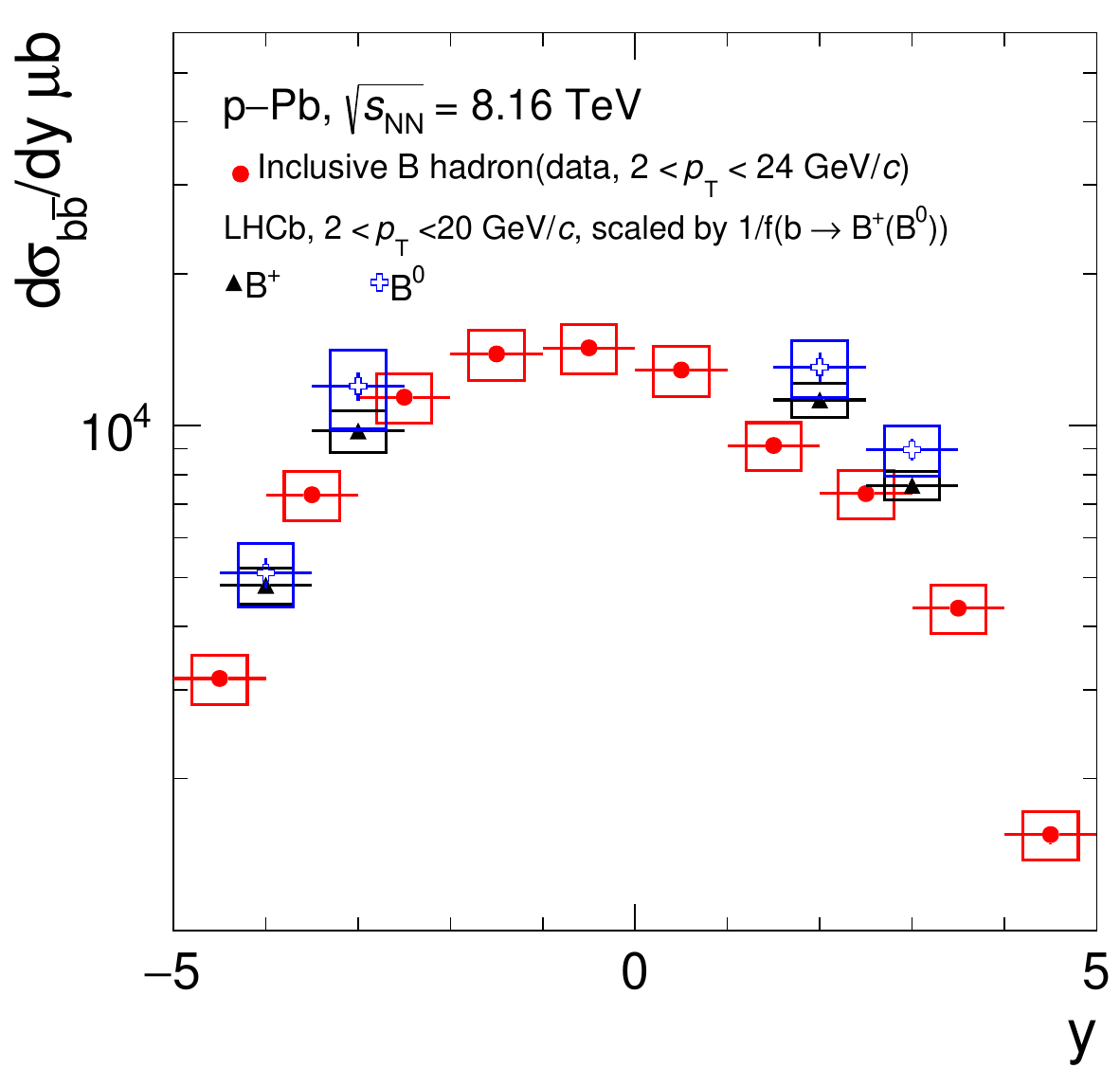}
        \caption{\y-differential unfolded results (red) scaled by production fractions and compared with LHCb measurements of \Bplus (black) and \Bzero (blue). The calculation covers $\pt = 2$--24~\GeVc and the measurements $\pt = 2$--20~\GeVc\cite{LHCb:2019avm}.}
 		\label{fig:expy}
 	\end{center}
 \end{figure}

The consistency between decay channels also constitutes a cross-check across collaborations. As a further validation, the unfolded inclusive beauty-hadron cross sections in \pPb collisions at \snn = 8.16~\TeV are scaled by the \Bplus and \Bzero  production fractions and compared with the corresponding exclusive LHCb measurements: the \pt-differential comparison in several rapidity intervals is shown in Fig.~\ref{fig:exppt} and the \y-differential one in Fig.~\ref{fig:expy}. The good agreement with the LHCb data within uncertainties provides additional validation of the unfolding procedure.

\subsection{Nuclear modification factors (\RpPb), mean $p_{T}$ square ($\langle p_{{T}}^{2} \rangle$) and total cross section in \pPb collisions at \snn = 5.02 and 8.16 TeV}

\begin{figure}[!htb]
 	\begin{center}
 		\includegraphics[width = 7.5cm]{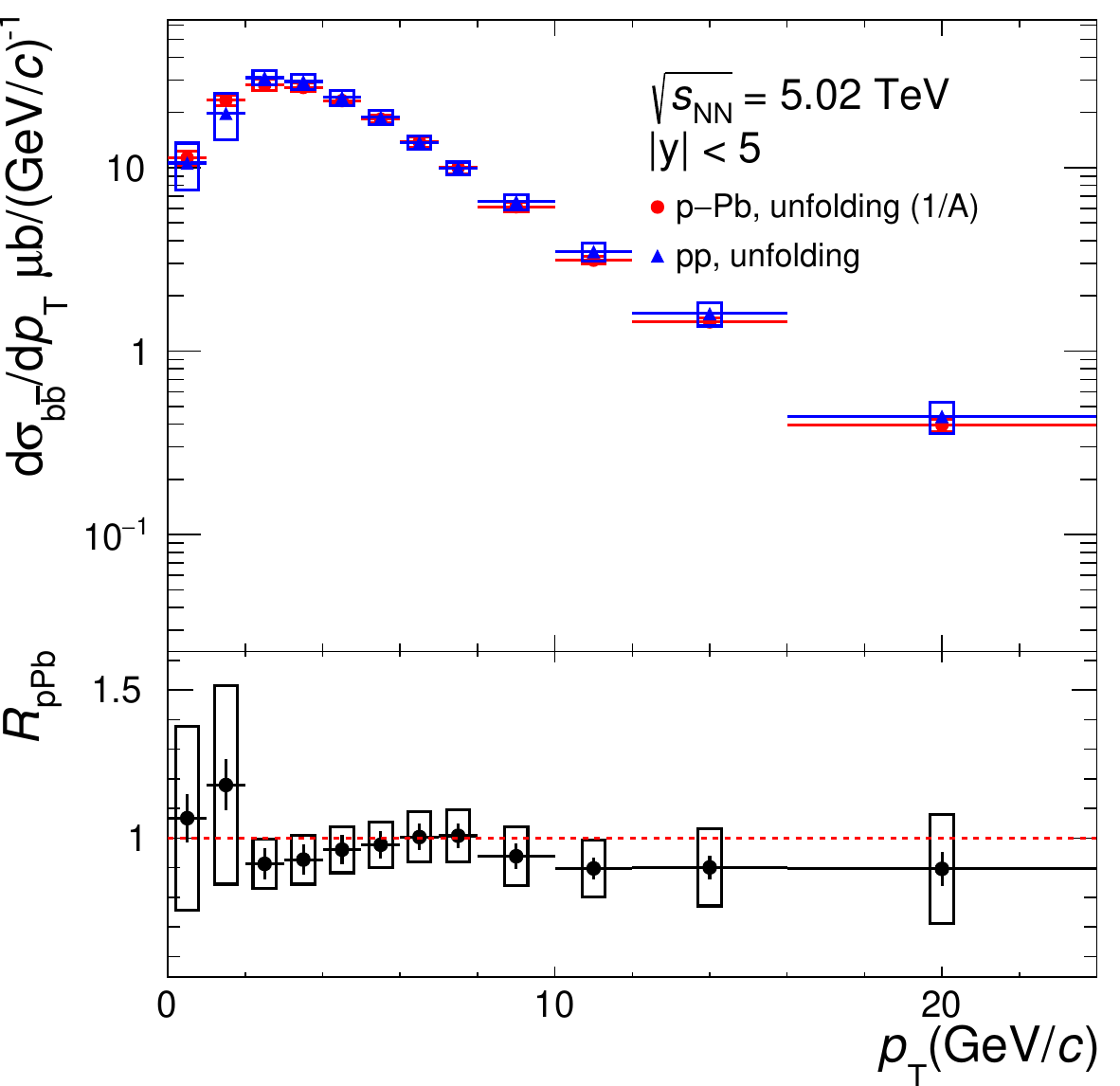}
            \includegraphics[width = 7.5cm]{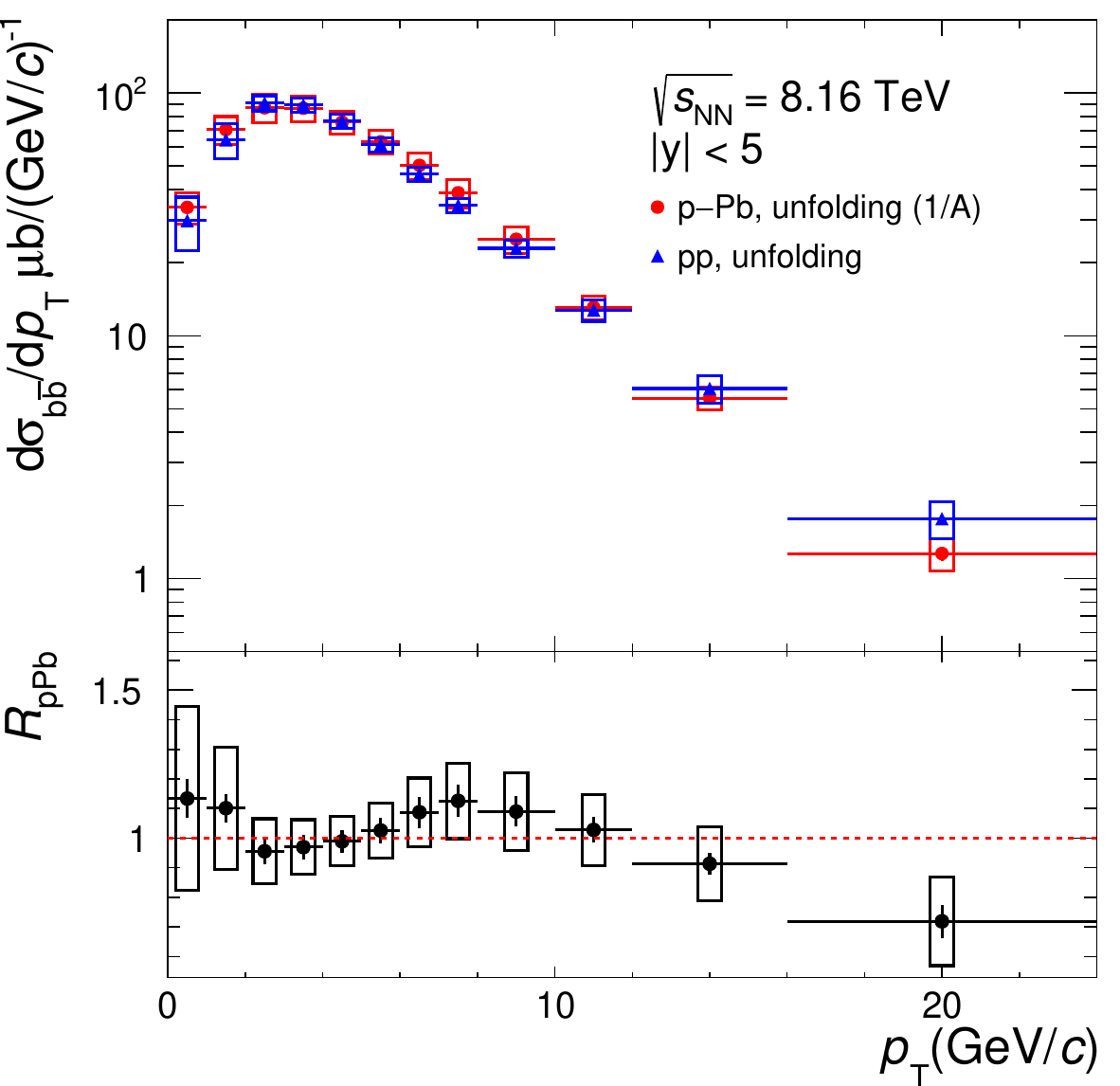}
 		\caption{Weighted-average \bbbar production cross section \dsigmadpt as a function of \pt in \pPb (scaled by $1/A$) and \pp collisions at \snn = 5.02~\TeV (left) and 8.16~\TeV (right) for $|y| < 5$. The nuclear modification factor \RpPb is shown in the lower panels.}
 		\label{fig:PtSpectra_combined_5020-ratio}
 	\end{center}
 \end{figure}

Figure~\ref{fig:PtSpectra_combined_5020-ratio} shows the comparison between the weighted-average \pt-differential \bbbar production cross sections in \pPb collisions (scaled by $1/A$) and the corresponding \pp reference cross sections. The nuclear modification factors \RpPb, defined as the ratio of these distributions, are shown in the lower panels.

\begin{figure}[!htb]
    \centering
    \includegraphics[width=7.5cm]
    {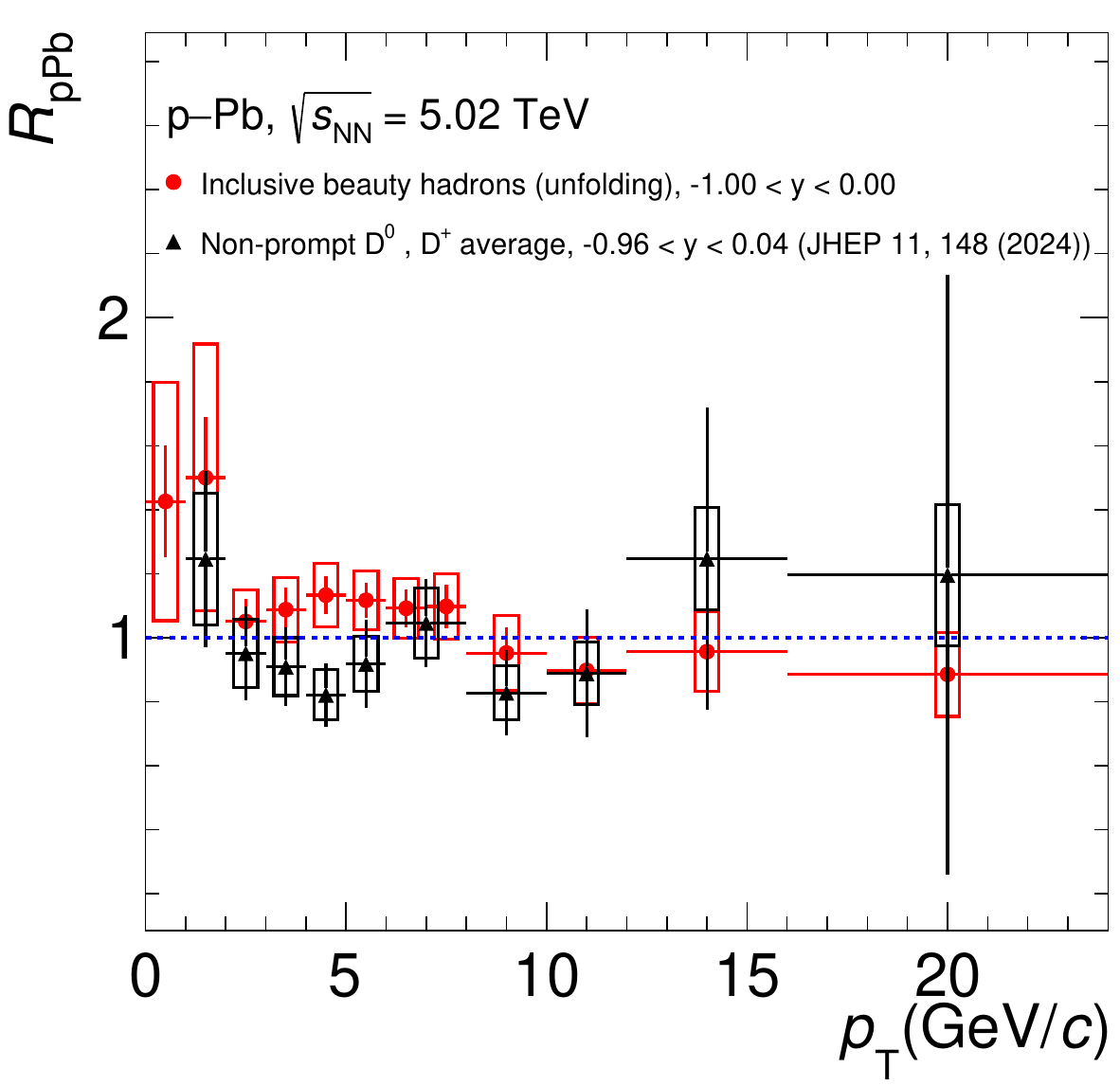}
    \includegraphics[width=7.5cm]
    {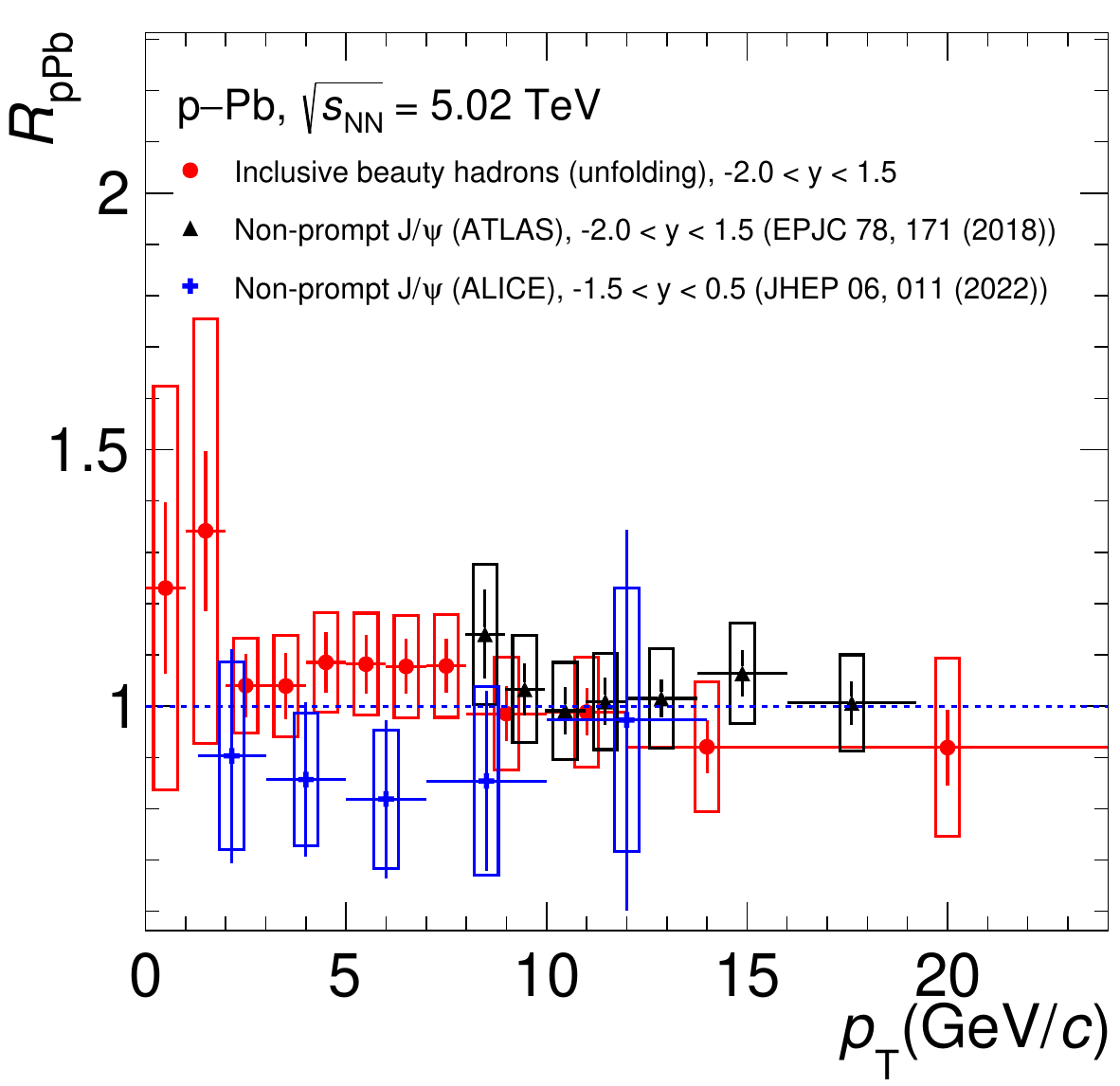}
    \caption{Nuclear modification factor \RpPb as a function of \pt in \pPb collisions at \snn = 5.02~\TeV. Left: unfolded results (red) compared with ALICE non-prompt D-meson data (black) for $-1.0 < y < 0.0$. Right: unfolded results (red) compared with ATLAS non-prompt \jpsi (black) and ALICE non-prompt \jpsi (blue) for $-2.0 < y < 1.5$ and $-1.5 < y < 0.5$, respectively.}
    \label{fig:RpPb_pt_5020}
\end{figure}

Figure~\ref{fig:RpPb_pt_5020} shows the nuclear modification factor as a function of \pt in several rapidity intervals in \pPb collisions at \snn = 5.02~\TeV. Since \RpPb is a ratio of \pPb to \pp cross sections obtained from the same decay channels, the branching-ratio uncertainties can be canceled. But considering the combined spectrum using different weights, the branch ratio canceled partially as following these formulas:
\begin{equation}
    P=w_{3}P_{D}+w_{4}P_{J,A}+w_{5}P_{J,L}; Q=w_{1}Q_{D}+w_{2}Q_{J}; R=P/Q\\
\end{equation}

\begin{equation}
    f_{D}^{pPb}=\frac{w_{3}P_{D}}{P}; f_{D}^{pp}=\frac{w_{1}Q_{D}}{Q}; f_{J}^{pPb}=\frac{w_{4}P_{J,A}+w_{5}P_{J,L}}{P};
    f_{J}^{pp}=\frac{w_{2}Q_{J}}{Q}\\
\end{equation}

\begin{equation}
    \sigma_{R,BR,D}=|R|r_{D}|f_{D}^{pPb}-f_{D}^{pp}|; \sigma_{R,BR,J}=|R|r_{J}|f_{J}^{pPb}-f_{J}^{pp}| 
\end{equation}
P and Q denote the combined p–Pb and pp cross sections, respectively. The subscripts D, J,A, and J,L denote the results obtained from the $D^{0}$
, ALICE J/$\psi$, and LHCb J/$\psi$ inputs. $w_{i}$ are the fixed combination weights, while $r_{D}$ and $r_{J}$ are the relative branching-fraction uncertainties.

\begin{figure}[!htb]
    \centering
    \includegraphics[width=7.5cm]
    {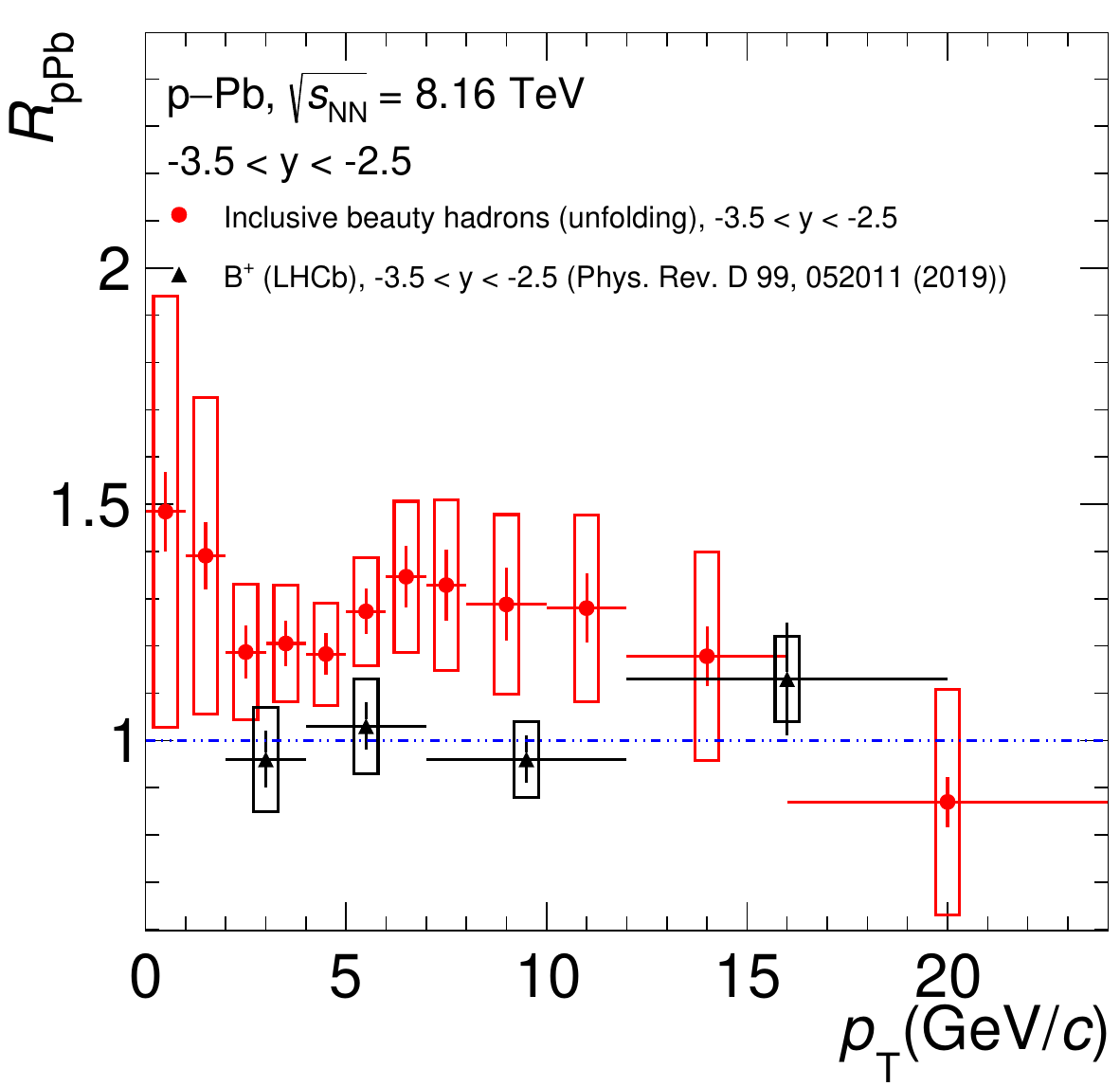}
    \includegraphics[width=7.5cm]
    {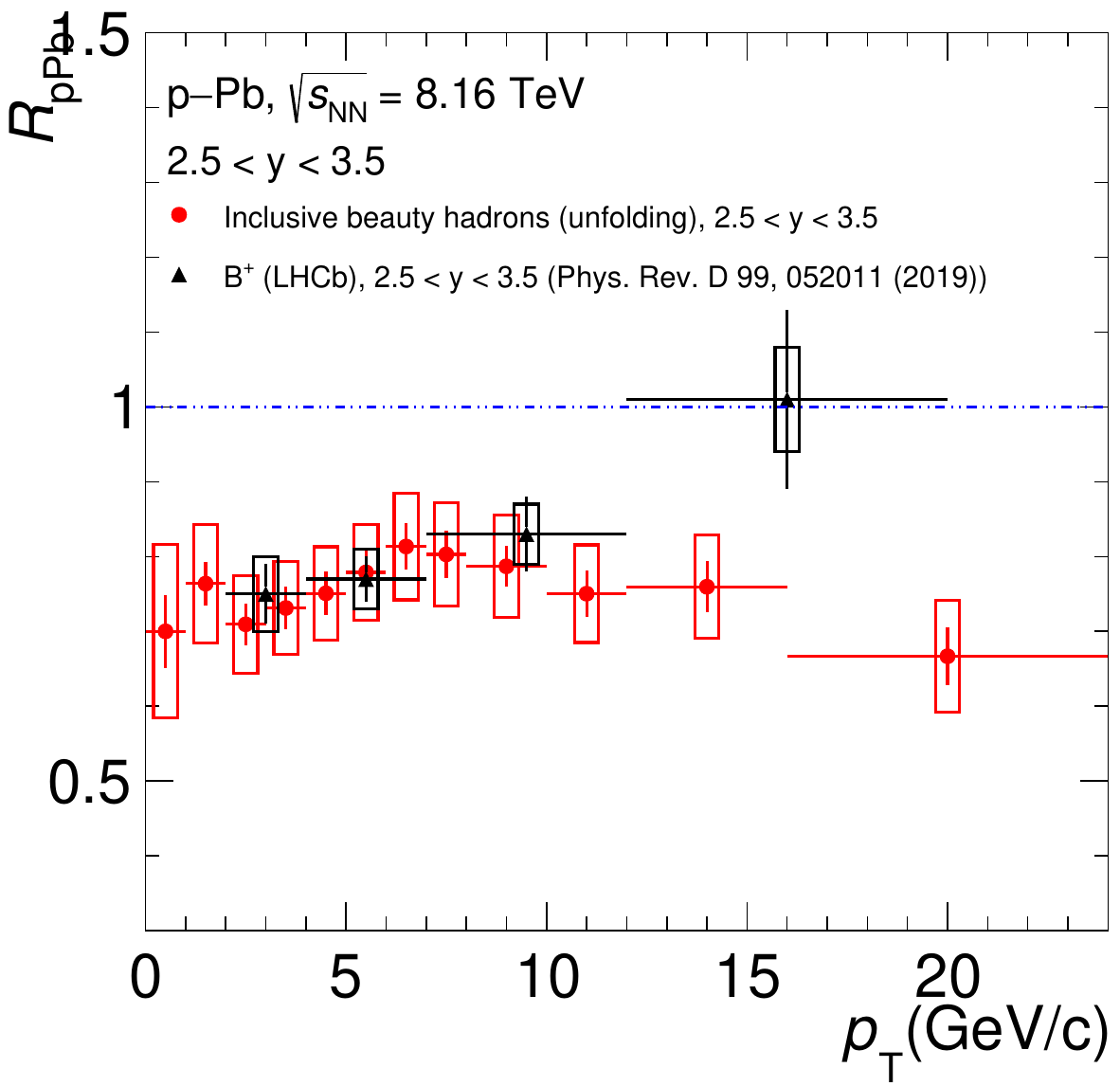}
    \caption{Nuclear modification factor \RpPb as a function of \pt in \pPb collisions at \snn = 8.16~\TeV for $-3.5 < y < -2.5$ (left) and $2.5 < y < 3.5$ (right). Unfolded results (black) are compared with LHCb \Bplus measurements (red).}
    \label{fig:RpPb_pt_8160}
\end{figure}

The unfolded results are compared with the ALICE measurement of non-prompt D mesons at midrapidity (left panel) and with the ATLAS and ALICE measurements of non-prompt \jpsi (right panel); all are consistent within uncertainties. The corresponding \pt-dependent \RpPb at \snn = 8.16~\TeV is shown in Fig.~\ref{fig:RpPb_pt_8160}, where the unfolded results agree with the LHCb \Bplus measurements within uncertainties.

\begin{figure}[!htb]
 	\begin{center}
 		\includegraphics[width = 7.5cm]{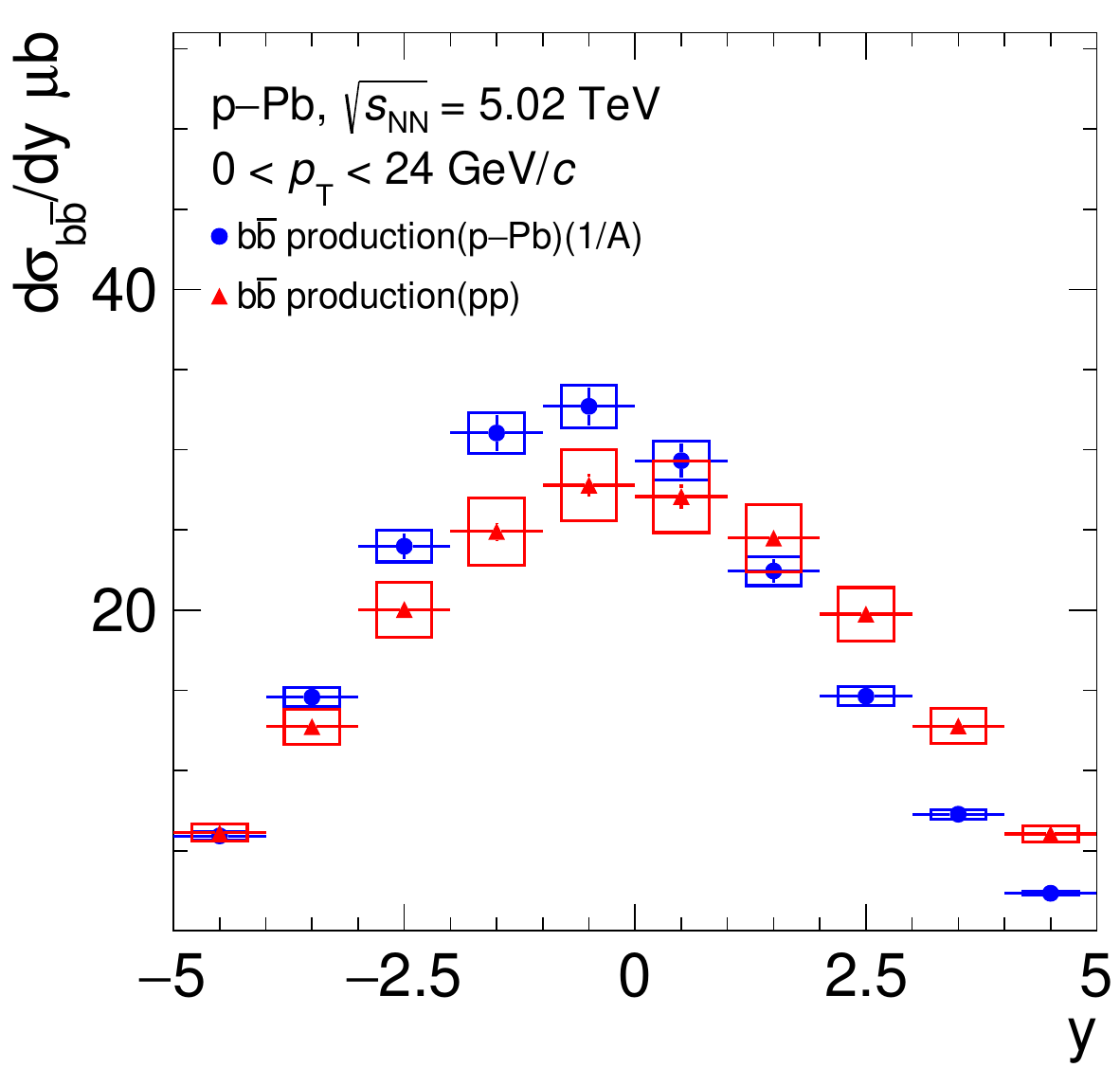}
            \includegraphics[width = 7.5cm]{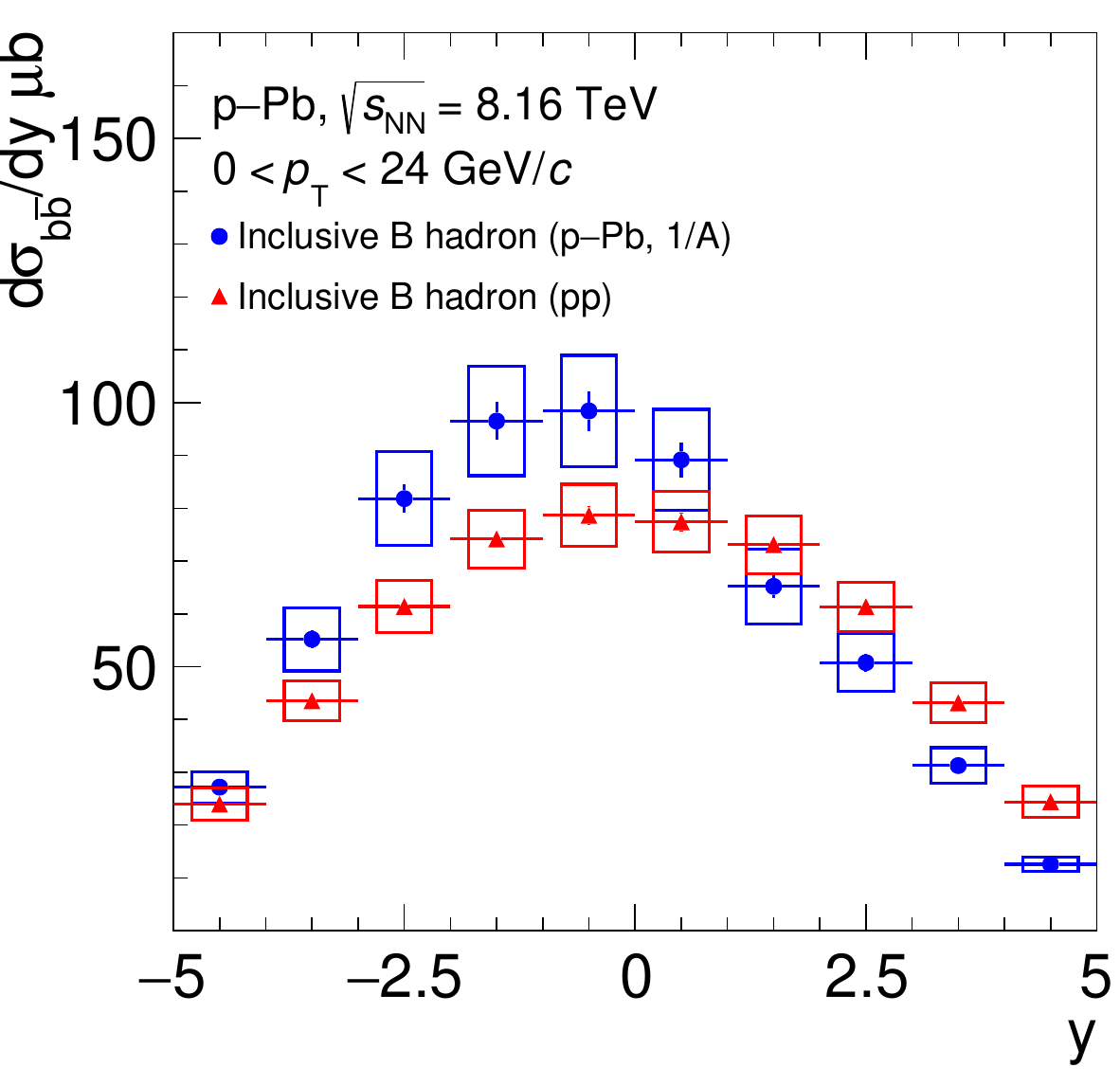
            }
 		\caption{\pt-integrated \bbbar cross section \dsigmady as a function of \y in \pPb (scaled by $1/A$) and \pp collisions at \snn = 5.02~\TeV (left) and 8.16~\TeV (right) for 0 $<$ \pt $<$ 24 \GeVc.}
 		\label{fig:Rapidity_combine_5020_8160}
 	\end{center}
 \end{figure}

The \pt-integrated \bbbar production cross sections \dsigmady as a function of \y in \pPb and \pp collisions at \snn = 5.02~\TeV (0 $<$ \pt $<$ 24) are compared in the left panel of Fig.~\ref{fig:Rapidity_combine_5020_8160}. The corresponding \RpPb as a function of rapidity is obtained from the ratio of the \pPb cross section, scaled by $1/A$, to the \pp reference. The scaled \pPb results follow a rapidity dependence qualitatively similar to that of \pp, with deviations that reflect cold-nuclear-matter effects. The corresponding comparison at \snn = 8.16~\TeV is shown in the right panel.

\begin{figure}[!htb]
 	\begin{center}
 		\includegraphics[width = 7.5cm]{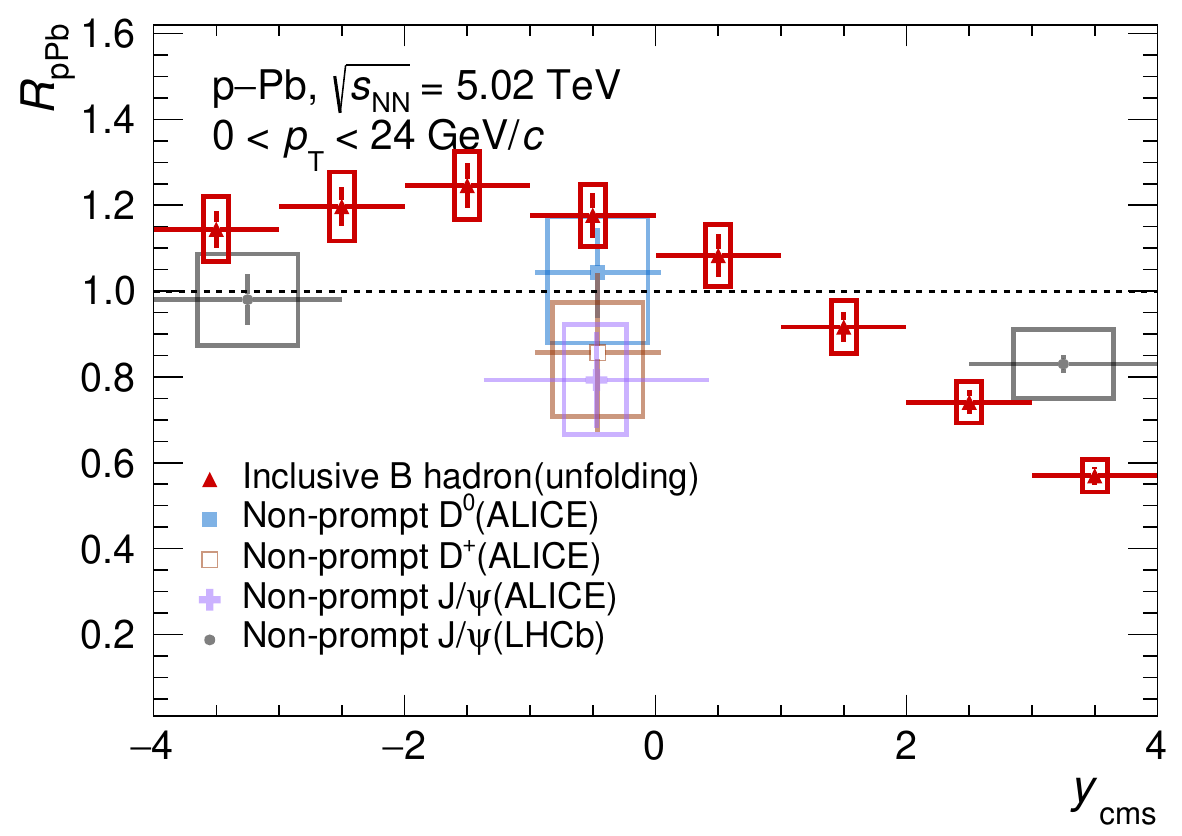}
            \includegraphics[width = 7.5cm]{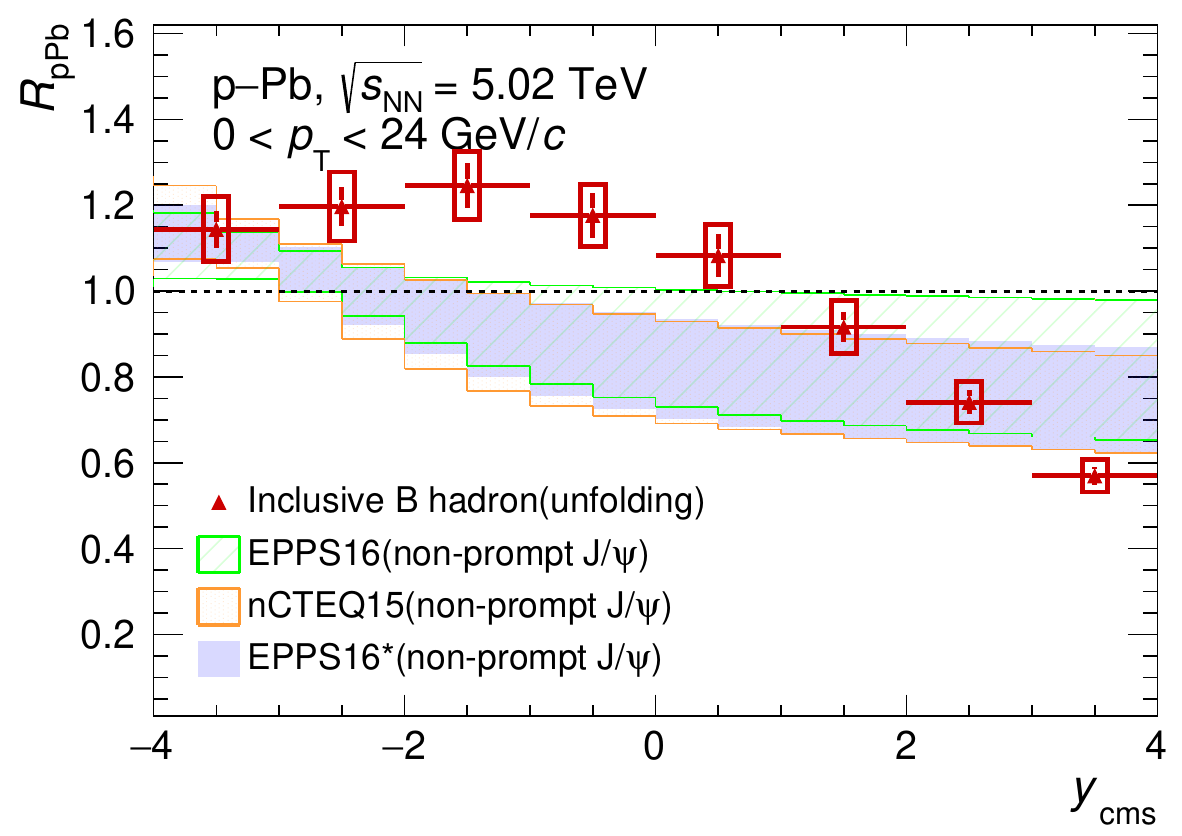}
 		\caption{Nuclear modification factor \RpPb as a function of \y at 0 $<$ \pt $<$ 24 \GeVc in \pPb collisions at \snn = 5.02~\TeV. Left: unfolded results (red) compared with ALICE and LHCb measurements of non-prompt \jpsi, \dzero, \dplus, and \Bplus. Right: unfolded results (red) compared with B-meson \RpPb calculations using the EPPS16, nCTEQ15, and EPPS16* nPDF sets.}
 		\label{fig:RpPb_Rapidity_5020}
 	\end{center}
 \end{figure}

Figure~\ref{fig:RpPb_Rapidity_5020} presents the nuclear modification factor at 0 $<$ \pt $<$ 24 \GeVc as a function of rapidity in \pPb collisions at \snn = 5.02~\TeV. The unfolded results are compared with experimental measurements from ALICE and LHCb and with calculations from the HELAC-onia generator using the EPPS16, nCTEQ15, and EPPS16* nPDF sets. The unfolded \RpPb is compatible with both within uncertainties. This measurement provides the first rapidity-differential beauty-hadron \RpPb covering the full rapidity range, connecting the midrapidity and forward-rapidity regions with a precision comparable to or better than that of existing measurements. 

\begin{figure}[!htb]
 	\begin{center}
 		\includegraphics[width = 7.5cm]{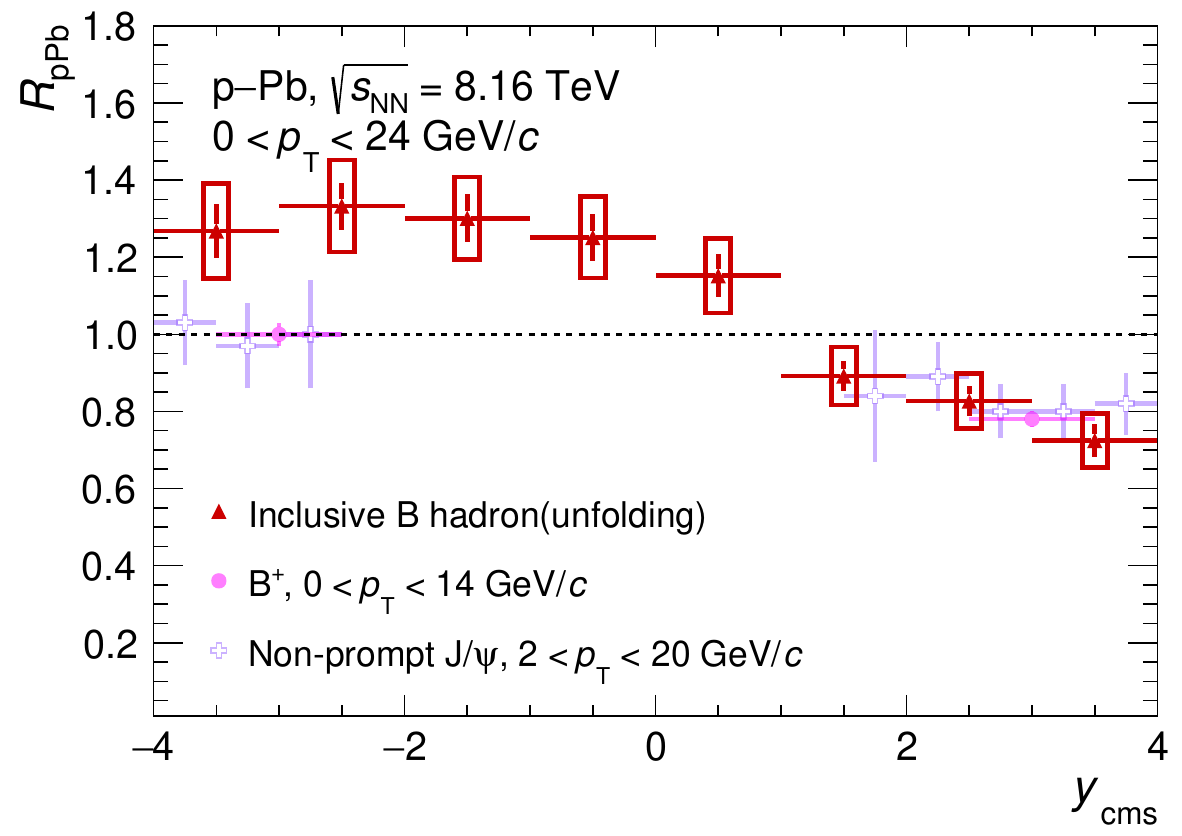}
            \includegraphics[width = 7.5cm]{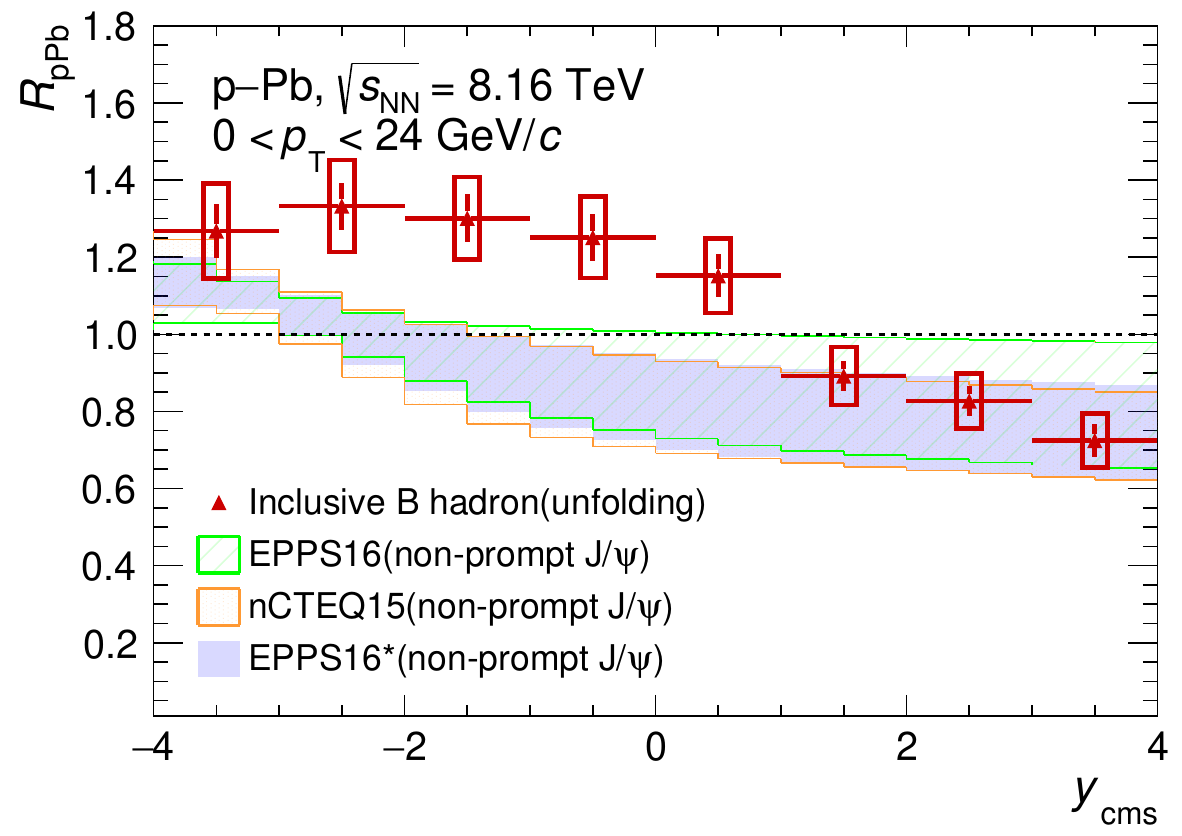}
 		\caption{Nuclear modification factor \RpPb as a function of \y at 0 $<$ \pt $<$ 24~\GeVc in \pPb collisions at \snn = 8.16~\TeV. Left: unfolded results (red) compared with LHCb measurements of non-prompt \jpsi and \Bplus. Right: unfolded results (red) compared with B-meson \RpPb calculations using the EPPS16, nCTEQ15, and EPPS16* nPDF sets.}
 		\label{fig:RpPb_Rapidity_8160}
 	\end{center}
 \end{figure}

The nuclear modification factor integrated over the measured \pt range ($\pt < 24$~\GeVc) as a function of rapidity in \pPb collisions at \snn = 8.16~\TeV is shown in Fig.~\ref{fig:RpPb_Rapidity_8160}. The unfolded results are compared with the LHCb measurements of non-prompt \jpsi and \Bplus mesons (left panel) and with model calculations employing the EPPS16, nCTEQ15, and EPPS16* nPDF sets (right panel). The unfolded \RpPb is consistent with both the experimental data and the theoretical predictions within uncertainties.
 
\begin{figure}[!htb]
 	\begin{center}
 		\includegraphics[width = 7.5cm]{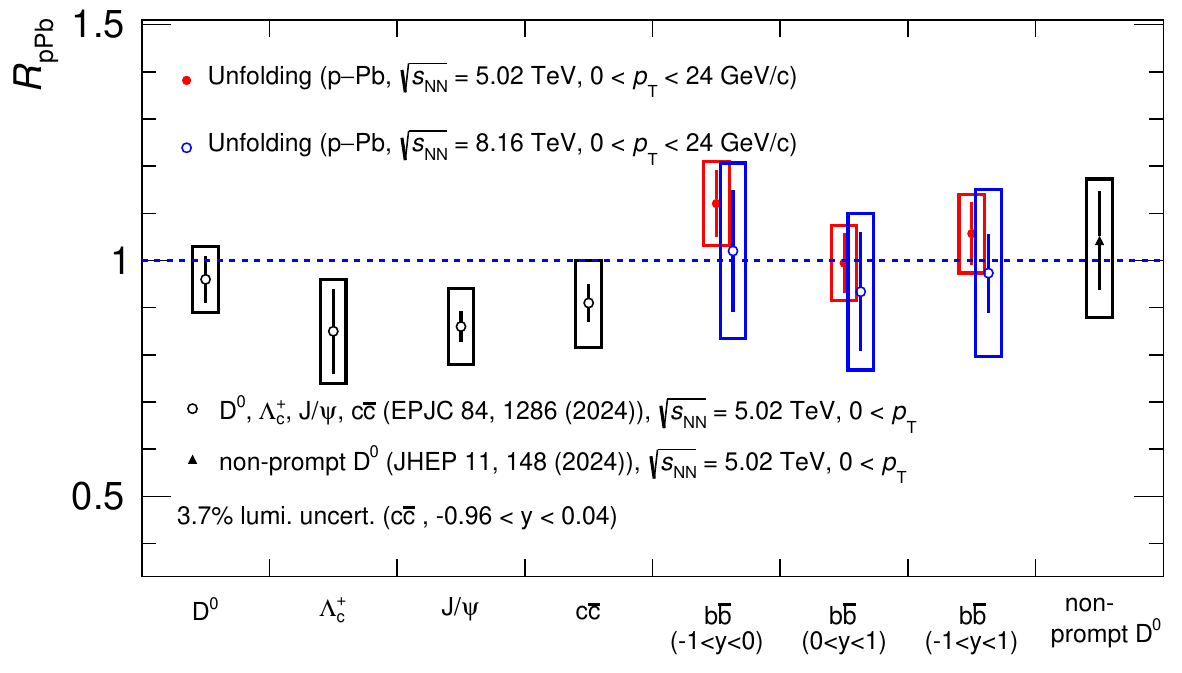}
 		\caption{Comparison of the nuclear modification factor of beauty hadrons (red) and charm hadrons (black) at midrapidity in \pPb collisions at \snn = 5.02~\TeV.}
 		\label{fig:RpPb_compare_ccbar_5020}
 	\end{center}
 \end{figure}

Figure~\ref{fig:RpPb_compare_ccbar_5020} compares the nuclear modification factor of beauty hadrons with that of charm hadrons at midrapidity in \pPb collisions at \snn = 5.02~\TeV. The beauty-hadron \RpPb is systematically larger than the charm-hadron one, indicating a weaker cold-nuclear-matter suppression for beauty, consistent with the expected mass hierarchy of CNM effects.

\begin{figure}[!htb]
 	\begin{center}
 		\includegraphics[width = 7.5cm]{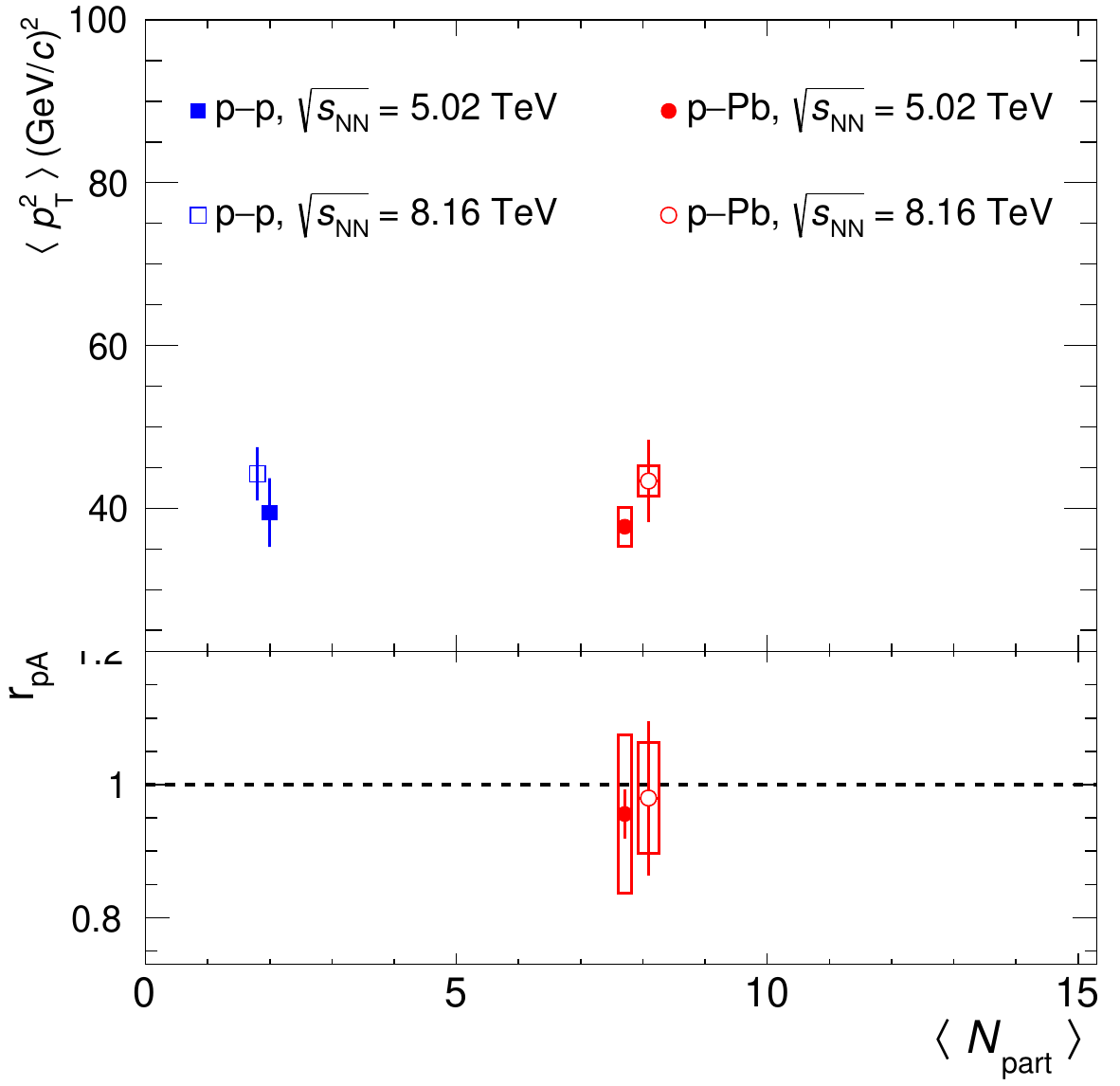}
 		\caption{Mean squared transverse momentum $\langle p_{{T}}^{2} \rangle$ of beauty hadrons as a function of $N_{{part}}$ in \pp collisions at $\sqrt{s} = 5.02$ and 8.16~\TeV and in \pPb collisions at \snn = 5.02 and 8.16~\TeV. The \pPb-to-\pp ratio is shown in the lower panel.}
 		\label{fig:mean_pt_5020}
 	\end{center}
 \end{figure}

The mean squared transverse momentum $\langle p_{{T}}^{2} \rangle$ of beauty hadrons in \pPb (red) and \pp (blue) collisions at \snn = 5.02 and 8.16~\TeV is shown in Fig.~\ref{fig:mean_pt_5020}. The $\langle N_{part} \rangle$ values are determined following the ALICE procedure described in Ref.~\cite{ALICE:2018centrality}. The upper panels display the individual values, the pp points are horizontally offset by 0.2 to distinguish the two
collision energies, and the lower panel shows the \pPb-to-\pp ratio. The $\langle p_{{T}}^{2} \rangle$ values are consistent between \pPb and \pp collisions at both energies within uncertainties, indicating no significant modification of the \pt-spectrum shape in \pPb collisions.

\FloatBarrier

The \bbbar production cross sections integrated over
$0 < \pt < 24~\GeV/c$ and $|y|<5$ in \pPb collisions at $\sqrt{s_{NN}}$ = 5.02 and 8.16~\TeV are

\begin{center}
    $\sigma_{\bbbar}^{\sqrt{s_{NN}}=5.02TeV} = 38.33 mb \pm 1.33 mb (stat.) \pm 2.40 mb (sys.)$
\end{center}

\begin{center}
    $\sigma_{\bbbar}^{\sqrt{s_{NN}}=8.16TeV} = 64.60 mb \pm 1.81 mb (stat.) \pm 5.87 mb (sys.)$
\end{center}

The result at \snn = 5.02~\TeV is consistent, within
uncertainties, with the full-phase-space estimate reported in
Ref.\cite{ALICE:2021lmn}, which was obtained by extrapolating the experimentally
measured visible cross section, although the two results correspond to slightly different phase-space definitions.

\section{Conclusions}
\label{Conclusions} 

The production cross sections of beauty hadrons in \pPb collisions at centre-of-mass energies of \snn = 5.02 and 8.16~\TeV have been measured via a data-driven Bayesian unfolding of non-prompt \dzero and \jpsi spectra, following the technique developed for \pp collisions in Ref.~\cite{Bai:2024pxk}. Compared with the corresponding \pp reference cross sections, these results provide a comprehensive characterisation of cold-nuclear-matter (CNM) effects on beauty production.

The unfolded \pt-differential and \pt-integrated \bbbar cross sections obtained from the ALICE non-prompt \dzero, ALICE non-prompt \jpsi, and LHCb non-prompt \jpsi channels are mutually consistent once unfolded to common rapidity and \pt ranges, and agree with directly reconstructed beauty-hadron measurements. This cross-channel consistency provides an important cross-check of the unfolding procedure and demonstrates the compatibility of the different experimental inputs. The final results are obtained as inverse-variance weighted averages of the individual channels.

The nuclear modification factor \RpPb has been determined from the ratio of the \pPb and \pp cross sections. The unfolded \RpPb values agree with existing experimental measurements and with nPDF-based calculations within uncertainties. This analysis provides a data-constrained determination of beauty-hadron \RpPb over a broad rapidity range, with a precision comparable to existing measurements in several rapidity intervals, providing complementary information for future global nPDF analyzes.The comparison suggests a weaker nuclear modification for beauty than for charm within the current uncertainties, which is qualitatively consistent with exceptions arising from their different mass scales.

The total \bbbar production cross sections in \pPb collisions at both energies have been reported, providing references for future measurements and model comparisons. Finally, the mean squared transverse momentum $\langle p_{{T}}^{2}\rangle$ of beauty hadrons is consistent between \pp and \pPb collisions at both energies, indicating no significant modification of the \pt-spectrum shape and indicating no significant modification of the mean squared transverse momentum within the current uncertainties.

\FloatBarrier

\section*{Acknowledgments}
\label{Acknowledgments}

This work was supported by the National Natural Science Foundation of China with Grant No.~12375141, No. 12475146 and No.~12675171, the Outstanding Leading Talent Team Program of Central China Normal University with Grant No. ~XJ2026000302, the Ministry of Science and Technology of China (MSTC) with Grant No.~2024YFA1610803.

\bibliographystyle{JHEP}
\bibliography{reference}

\end{document}